\documentclass[twocolumn, pra, superscriptaddress]{revtex4-2}
\pdfoutput=1
\usepackage{lmodern}
\usepackage[utf8]{inputenc}
\usepackage[english]{babel}
\usepackage[T1]{fontenc}
\usepackage{amsmath,amsthm,amssymb}
\usepackage{mathtools}
\usepackage[colorlinks=true]{hyperref}
\usepackage{adjustbox}
\usepackage{multirow}
\usepackage{stackrel}
\usepackage{graphicx}
\usepackage{cancel}
\usepackage[all]{hypcap}
\usepackage{comment}
\usepackage{tabularx}
\usepackage{color}
\usepackage{braket}
\usepackage{ragged2e}

\newcommand{\be}{\begin{equation}}
\newcommand{\ee}{\end{equation}}
\newcommand{\ba}{\begin{eqnarray}}
\newcommand{\ea}{\end{eqnarray}}

\newcommand{\mycomment}[1]{}
\newcommand{\addtuc}{School of Electrical and Computer Engineering, Technical University of Crete, Chania, Greece 73100}

\newcommand{\addcqt}{Centre for Quantum Technologies, National University of Singapore, 3 Science Drive 2, Singapore 117543}

\newcommand{\addihpc}{Institute of High Performance Computing, Agency for Science, Technology \& Research (A*STAR), 1 Fusionopolis Way, \#16-16 Connexis, Singapore 138632}

\newcommand{\addangel}{AngelQ Quantum Computing, 531A Upper Cross Street, \#04-95 Hong Lim Complex, Singapore 051531}

\begin{document}

\title{Shallow quantum circuits for efficient preparation of Slater determinants and correlated states on a quantum computer}
% \date{\today}

\author{Chong Hian Chee}
\email{ch.chee@u.nus.edu}
\affiliation{\addcqt}

\author{Daniel Leykam}
\affiliation{\addcqt}

\author{Adrian M. Mak}
\affiliation{\addihpc}

\author{Dimitris G. Angelakis}
\email{dimitris.angelakis@gmail.com}
\affiliation{\addcqt}
\affiliation{\addtuc}
\affiliation{\addangel}
\date{\today}
\begin{abstract}
Fermionic ansatz state preparation is a critical subroutine in many quantum algorithms such as Variational Quantum Eigensolver for quantum chemistry and condensed matter applications. The shallowest circuit depth needed to prepare Slater determinants and correlated states to date scale at least linearly with respect to the system size $N$. Inspired by data-loading circuits developed for quantum machine learning, we propose an alternate paradigm that provides shallower, yet scalable ${\mathcal{O}}(d \log_2^2N)$ two-qubit gate depth circuits to prepare such states with d-fermions, offering a subexponential reduction in $N$ over existing approaches in second quantization, enabling high-accuracy studies of $d{\ll}{\mathcal{O}}{\left(N / \log_2^2 N\right)}$ fermionic systems with larger basis sets on near-term quantum devices.
\end{abstract}

\maketitle
\section{Introduction}
Quantum computers promise the ability to solve hard many-body problems in quantum chemistry and condensed matter physics, including computation of ground state energies and simulation of quantum dynamics~\cite{bauerQuantumAlgorithmsQuantum2020, mottaEmergingQuantumComputing2022, daleyPracticalQuantumAdvantage2022}. The relevant quantum algorithms frequently involve quantum state preparation as a key step. For example, the success probability of quantum phase estimation is determined by the overlap of a trial ansatz state with the eigenstate of interest~\cite{abramsSimulationManyBodyFermi1997, abramsQuantumAlgorithmProviding1999, aspuru-guzikSimulatedQuantumComputation2005}. Thus, efficient preparation of high quality ansatz states is crucial for many-body applications of quantum computing~\cite{leeThereEvidenceExponential2022,bhartiNoisyIntermediatescaleQuantum2022}.

Most existing methods for preparing fermionic ansatzes use second quantization with Jordan-Wigner mapping~\cite{jordanUeberPaulischeAequivalenzverbot1928} to efficiently represent the quantum many-body fermionic wavefunction using a number of qubits that scales linearly in the system size~\cite{mottaEmergingQuantumComputing2022,bauerQuantumAlgorithmsQuantum2020}. Widely-used fermionic ansatzes typically fall into two broad classes: The first class consists of hardware-efficient ansatzes which use parameterized hardware-native gates to minimize the depth of the quantum circuit~\cite{kandalaHardwareefficientVariationalQuantum2017}, but are difficult to optimize~\cite{bittelTrainingVariationalQuantum2021} and do not guarantee an accurate representation of the desired quantum state~\cite{tillyVariationalQuantumEigensolver2022}. The second class consists of problem-inspired ansatzes which are more promising and explicitly incorporate the physics of the system of interest, but require deeper circuits that scale polynomially in system size, which exacerbates errors due to quantum noise and decoherence~\cite{weckerSolvingStronglyCorrelated2015, kivlichanQuantumSimulationElectronic2018, aruteHartreeFockSuperconductingQubit2020, anandQuantumComputingView2022, evangelistaExactParameterizationFermionic2019, wangResourceOptimizedFermionicLocalHamiltonian2021, kottmannOptimizedLowdepthQuantum2022, tangQubitADAPTVQEAdaptiveAlgorithm2021}, limiting state-of-the-art demonstrations to less than a hundred qubits~\cite{obrienPurificationbasedQuantumError2022, tazhigulovSimulatingModelsChallenging2022}, leaving studies of chemically-relevant molecular systems requiring more than $10^{2}$--$10^{3}$ qubits well out of reach~\cite{elfvingHowWillQuantum2020,nagyBasisSetsQuantum2017}.

\begin{figure}
\centering
\includegraphics[width=0.9\columnwidth, page=1]{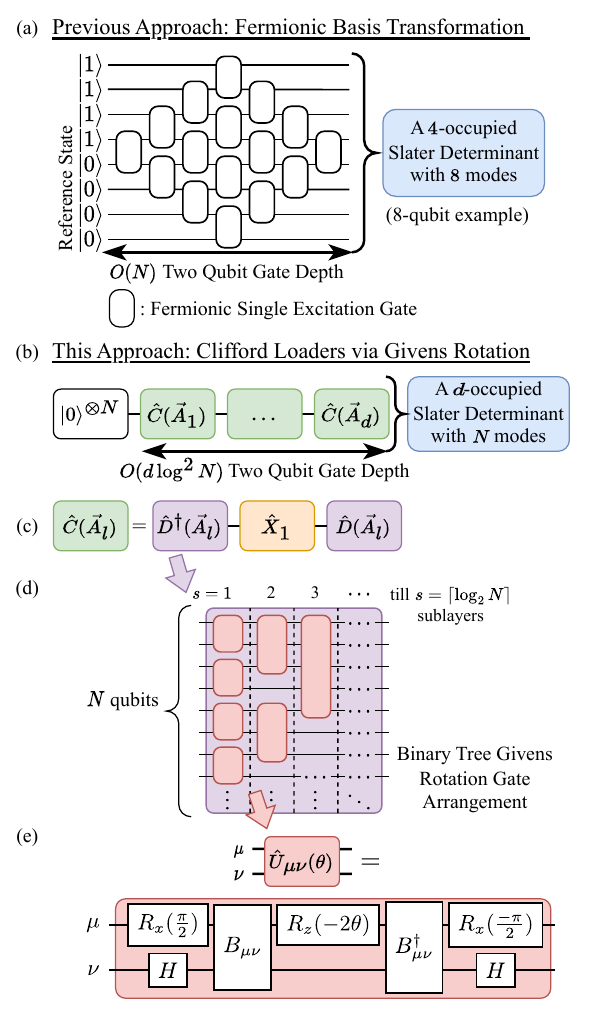}
\caption{Different approaches for preparing $d$ occupied Slater determinant of $N$ modes on a quantum computer, assuming Jordan-Wigner mapping. (a) Existing approaches use a linear-depth mesh of fermionic single excitation gates to apply a fermionic basis transformation to a reference state~\cite{ weckerSolvingStronglyCorrelated2015, kivlichanQuantumSimulationElectronic2018, aruteHartreeFockSuperconductingQubit2020}. (b) The proposed Clifford loaders via the Givens rotation approach applies a sequence $\hat{C}$ gates $d$ times to an all-zero state of $N$ qubits, $|0\rangle^{\otimes N}$. (c) $\hat{C}$ consists of two products of multiple Givens rotations $\hat{D}$, $\hat{D}^\dagger$ that sandwich a Pauli-X gate on the first qubit. (d) The Givens rotations $\hat{U}_{\mu\nu}$ are arranged in a binary tree. (e) $\hat{U}_{\mu\nu}$ is decomposed using Pauli rotation gates $R_x$, $R_z$ acting on qubit $\mu$, Hadamards $H$ acting on qubit $\nu$, and CNOT ladders $B_{\mu\nu}$ acting on all qubits between $\mu$ and $\nu$. $\theta$ and $\vec{A}$ are scalar and vector parameters respectively.}
\label{fig1}
\end{figure}

% mean-field hartree fock states, slater determinants
The shallowest general-purpose problem-inspired ansatz states to date are mean-field Hartree-Fock states~\cite{weckerSolvingStronglyCorrelated2015, kivlichanQuantumSimulationElectronic2018, aruteHartreeFockSuperconductingQubit2020}, which are Slater determinants that can be prepared using a mesh of fermionic single-excitation gates which have a linear $\mathcal{O}(N)$ two-qubit gate depth in the number of qubits $N$, as shown in Fig.~\ref{fig1}(a). While Hartree-Fock states are efficiently simulatable using classical computers, they nevertheless serve as a useful starting point for quantum computers to prepare more interesting classically-intractable correlated quantum ansatzes, such as the unitary coupled cluster ansatz, which incorporates quantum correlations by applying number-conserving multi-fermion excitation operators to a reference Hartree-Fock state~\cite{anandQuantumComputingView2022, romeroStrategiesQuantumComputing2018, evangelistaExactParameterizationFermionic2019, wangResourceOptimizedFermionicLocalHamiltonian2021, kottmannOptimizedLowdepthQuantum2022, tangQubitADAPTVQEAdaptiveAlgorithm2021}.

% Givens rotation 
Fermionic excitation operators are examples of Givens rotations gates which perform rotations in a two-dimensional fermionic subspace of a larger Hilbert space and together with its controlled-variants form a universal quantum gate set to realize any particle conserving unitaries~\cite{arrazolaUniversalQuantumCircuits2022, anselmettiLocalExpressiveQuantumnumberpreserving2021}. Therefore, such Givens rotations gates have been helpful for preparing various fermionic states in quantum chemistry and condensed matter applications~\cite{weckerSolvingStronglyCorrelated2015, anandQuantumComputingView2022, aruteHartreeFockSuperconductingQubit2020, kivlichanQuantumSimulationElectronic2018, evangelistaExactParameterizationFermionic2019, yordanovEfficientQuantumCircuits2020, magoulasCNOTEfficientCircuitsArbitrary2023, cheeComputingElectronicCorrelation2022}. Recently, such gates have also attracted interest in the context of quantum linear algebra, where they were used to construct shallow-depth ``Clifford loaders'' gates, which are linear combinations of anti-commuting operators, as a means to efficiently encode $d$-dimensional subspaces of $\mathbb{R}^N$ into a $N$-qubit state~\cite{kerenidisQuantumMachineLearning2022, stewartComputingTheCSDecomposition1982, gawlikBackwardStableAlgorithm2018, johriNearestCentroidClassification2021}, enabling potential end-to-end quantum speedups for several quantum machine learning and linear algebra problems including determinant sampling and topological data analysis~\cite{kerenidisQuantumMachineLearning2022}. It is thus timely to consider whether the implementation of Clifford loaders via the Givens rotation are useful for preparing of fermionic ansatz states.

% Our work
Here we shall show how a $d$-fermion Slater determinant can be prepared in second quantization using Clifford loaders with an overall $\mathcal{O}(d\log_2 N)$ Givens rotation gate depth. An overall $\mathcal{O}(d\log_2^2 N)$ two-qubit gate depth scaling is thus achieved using Givens rotation gates with $\mathcal{O}(\log_2 N)$ two-qubit gate depth under Jordan-Wigner fermion-to-qubit mapping. This is more depth-efficient than existing approaches that scale as $\mathcal{O}(N)$ in two-qubit gate depth, with respect to the system size $N$ for a sufficiently slow growing $d$. We also show how this approach can be extended using the same preparation technique for Slater determinants to prepare fermionic $L$-wise correlated ansatz states, that correlates between $L$-tuples of fermionic modes in the ansatz, to yield quantum circuits that is shallower than that of Slater determinant by at least a factor of $L$. Finally, to validate our correlated ansatz, we shall demonstrate how the $L{=}2$ pairwise correlated ansatz can be used to capture a significant fraction of correlation energy using an example of hydrogen chains up to $N{=}20$ qubits, where pairwise electronic correlation is likely to be significant. Our results establish Clifford loaders constructed out of Givens rotations as a promising method for efficient, practical, and scalable preparation of fermionic ansatz states for large quantum chemistry applications on near-term quantum computers.

\section{Preparing Slater determinant using Shallow Circuits}
We begin by showing how a Slater determinant with $d$ occupied and $N-d$ unoccupied fermionic modes can be prepared using a shallow quantum circuit. An arbitrary Slater determinant $|\Psi_{1}\rangle$ is defined as~\cite{weckerSolvingStronglyCorrelated2015, wanMatchgateShadowsFermionic2022,ortizQuantumAlgorithmsFermionic2001,jiangQuantumAlgorithmsSimulate2018}
\be
|\Psi_{1}(A)\rangle \vcentcolon= \prod_{l=1}^{d}\sum^N_{\mu=1} A_{\mu l} \hat{a}^\dagger_\mu  |\textrm{vac}\rangle, \label{eq:slater}
\ee
\noindent
where $A$ is an $N{\times}d$ real matrix such that all $d$ columns are orthogonal and normalized, $|\textrm{vac}\rangle$ is a vacuum state, and $\hat{a}^\dagger_\mu$ is a creation operator acting on the $\mu$\textsuperscript{th} mode. While the definition in Eq.~\eqref{eq:slater} is pedagogically convenient, it requires non-unitary operators $\Sigma^N_{\mu=1} A_{\mu l} \hat{a}^\dagger_\mu$ which cannot be directly implemented on a quantum circuit. Consequently, the most efficient method to prepare Slater determinants to date was to perform a fermionic basis transformation to a reference Slater determinant state for a given skew-Hermitian parameter matrix $\kappa$ as 
\be
|\Psi_{1}(A)\rangle=\exp\left[\sum_{\mu,\nu=1}^N\kappa_{\mu\nu}\hat{a}^\dagger_\mu\hat{a}_\nu\right]\prod_{r=1}^{d} a^\dagger_r |\textrm{vac}\rangle,
\ee
\noindent 
where the fermionic basis transformation is implemented as a linear-depth mesh of fermionic single-excitation gates $\exp[\theta(\hat{a}^\dagger_\mu\hat{a}_{\nu}-\hat{a}^\dagger_\nu\hat{a}_{\mu})]$, as shown in Fig.~\ref{fig1}(a)~\cite{aruteHartreeFockSuperconductingQubit2020,kivlichanQuantumSimulationElectronic2018}.

% Use an alternate definition
We propose to improve the circuit depth efficiency of preparing Slater determinants using the equivalent form
\be
|\Psi_{1}(A)\rangle\vcentcolon= \prod_{l=1}^{d}\sum_{\mu=1}^{N}A_{\mu l}\hat{p}_\mu |\textrm{vac}\rangle, \label{eq:pk}
\ee
\noindent
as shown in Appendix~\ref{apx:equivalence_slater}, where we use anti-commuting operators 
\be
\hat{p}_\mu=\hat{a}^\dagger_\mu+\hat{a}_\mu, \label{eq:anti_comm_slater}
\ee
\noindent
with the relation $\{\hat{p}_\mu,\hat{p}_\nu\}{=}2\delta_{\mu\nu}\mathbf{I}$ as shown in Appendix~\ref{apx:anti_comm}. Using the anti-commuting operators $\hat{p}_\mu$ instead of $\hat{a}^\dagger_\mu$ allows us to exploit the recent result in Ref.~\cite{kerenidisQuantumMachineLearning2022} that provides a shallow $\mathcal{O}(d \log_2 N)$ Givens rotation gate depth decomposition of the Clifford loader
\be
\hat{C}(\vec{A}_l) = \sum^{N}_{\mu=1} A_{\mu l}\hat{p}_\mu, \label{eq:cliff_load_def}
\ee
\noindent
for some normalized column $\vec{A}_l$. Applying the Clifford loaders $\hat{C}(\vec{A}_l)$ in succession $d$ times on a vacuum state $|\textrm{vac}\rangle$, each with orthogonal columns $\vec{A_l}$, $l{=}1,\ldots,d$ from matrix $A$ generates the desired Slater determinant
\be
|\Psi_{1}(A)\rangle = \prod_{l=1}^{d} \hat{C}(\vec{A}_l)  |\textrm{vac}\rangle. \label{eq:slater_cliff}
\ee
\noindent
as shown in Fig.~\ref{fig1}(b). In the case of $d{>}\frac{N}{2}$, we may apply the Clifford loader $N{-}d$ times instead, followed by occupation-vacant mode swap to all Fock basis at the end, that is equivalent to a Pauli-X bitflip to all qubits under the Jordan-Wigner mapping. The Slater determinant in Eq.~\eqref{eq:slater_cliff} can be simplified to~\cite{kerenidisQuantumMachineLearning2022, vourdasExteriorCalculusFermionic2018} 
\be
|\Psi_{1}(A)\rangle = \sum_{|B|=d } \textrm{det}(A_B) |B\rangle, \label{eq:slater_subspace}
\ee
\noindent
where the sum is over all possible combinations of the ordered set $B$ containing $d$-unique integers from $1$ to $N$, $A_B$ is a $d{\times}d$ matrix minor of $A$ whose row indexes are restricted by $B$, and $|B\rangle$ denotes an $N$-mode Fock basis with occupied modes indexed by $B$ as shown in Appendix~\ref{apx:cliff_load}.

The Clifford loader $\hat{C}(\vec{A}_l)$ in Eq.~\eqref{eq:cliff_load_def} can be decomposed as
\be
\hat{C}(\vec{A}_l)=\hat{D}(\vec{A}_l)\hat{p}_1\hat{D}^\dagger(\vec{A}_l),  \label{eq:cliff_implement}
\ee
\noindent
as shown in Fig.~\ref{fig1}(c), where the operator $\hat{p}_1{=}\hat{a}^\dagger_1+\hat{a}_1$ acting on the first mode is sandwiched by two products of multiple Givens rotations, termed elsewhere as unary data loaders $\hat{D}$.~\cite{kerenidisQuantumMachineLearning2022, kerenidisClassicalQuantumAlgorithms2021, johriNearestCentroidClassification2021} . Its conjugate transpose $\hat{D}^\dagger$ is expressed as
\be
\hat{D}^\dagger(\vec{A}_l)=\prod_{s=1}^{\lceil\log_2 N\rceil}\left[\prod_{\mu,\nu\in \mathfrak{T}_s}\hat{U}_{\mu\nu}(\theta^{(sl)}_{\mu\nu})\right], \label{eq:unary_data_loader}
\ee
\noindent
where the Givens rotations
\be
\hat{U}_{\mu\nu}(\theta)=\exp[\theta\hat{p}_\mu\hat{p}_\nu] \label{eq:givens_slater}
\ee
\noindent
are arranged in a binary tree pattern according to the set of $(\mu,\nu)$ indexes $\mathfrak{T}_s=\{(\mu,\nu)|\mu=2^s(k{-}1){+}1,\,\nu=2^{s{-}1}(2k{-}1){+}1,\,k\in\mathbb{Z}^+\backslash 0\}$ for each sublayer $s{\in}\{1,\ldots,\lceil\log_2 N\rceil\}$ as shown in Fig.~\ref{fig1}(d). 

By treating $\vec{A}_l{=}(A_{1l},{\ldots},A_{Nl})$ in Eq.~\eqref{eq:unary_data_loader} as a vector in the basis of $\{\hat{p}_1,\ldots,\hat{p}_N\}$, we exploit the Givens rotations property that
\be
\hat{U}_{\mu\nu}(\theta)\hat{p}_r\hat{U}^{\dagger}_{\mu\nu}(\theta)=\begin{dcases} \cos(2\theta)\hat{p}_r+ \sin(2\theta)\hat{p}_\nu &r = \mu, \\
\cos(2\theta)\hat{p}_r -\sin(2\theta) \hat{p}_\mu&r = \nu, \\
\hat{p}_r &r \ne \mu,\nu,\end{dcases}
\ee
\noindent
to obtain the required rotation angles $\theta^{(sl)}_{\mu\nu}{=}\frac{1}{2}\arctan\frac{A^s_{\nu l}}{A^s_{\mu l}}$ classically by numerically performing parallel Givens rotations on $\vec{A}_l$ that correspond to the sequence in $\hat{D}^\dagger$ that successively zeros out the vector elements till the first element of $\vec{A}_l$  becomes $A_{1l}{=}1$ corresponding to $\hat{p}_1$. 

\begin{figure}
\centering
\includegraphics[width=0.7\columnwidth, page=1]{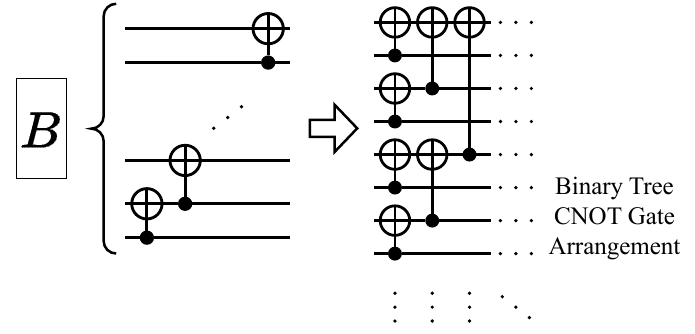}
\caption{Linear-depth cascading CNOT ladder $B$ in the decomposition of Givens rotations gate $\hat{U}$ is replaced by a non-equivalent logarithmic-depth circuit that has no effect on $\hat{U}$.}
\label{fig2}
\end{figure}

Under the Jordan-Wigner mapping, the Givens rotation in Eq.~\eqref{eq:givens_slater} maps to $\exp[-i\theta\hat{Y}_\mu\hat{X}_\nu{\otimes_{r=\mu+1}^{\nu-1}}\hat{Z}_r]$, which is a Pauli string rotation gate that can be easily implemented on a quantum circuit~\cite{whitfieldSimulationElectronicStructure2011,yordanovEfficientQuantumCircuits2020} as shown in Fig.~\ref{fig1}(e). The CNOT ladder $B_{\mu\nu}$ in the gate decomposition of the Givens rotation $\hat{U}_{\mu\nu}$ serves the purpose of encoding the parity of non-exciting qubits into the rotation gate and it consists of a cascade of CNOT gate~\cite{whitfieldSimulationElectronicStructure2011, yordanovEfficientQuantumCircuits2020}. However, this CNOT ladder can be replaced by a non-equivalent binary tree CNOT gate arrangement~\cite{kerenidisQuantumMachineLearning2022} shown in Fig.~\ref{fig2}, without any effect on the Givens rotation $\hat{U}_{\mu\nu}$, thereby reducing the CNOT depth from linear to logarithmic in $N$. Thus, by implementing $d$ such Clifford loaders with these Givens rotation gates, it enables us to prepare Slater determinants with a shallow $\mathcal{O}(d\log^2_2 N)$ two-qubit gate depth quantum circuits.

\begin{figure}
\centering
\includegraphics[width=1\columnwidth, page=1]{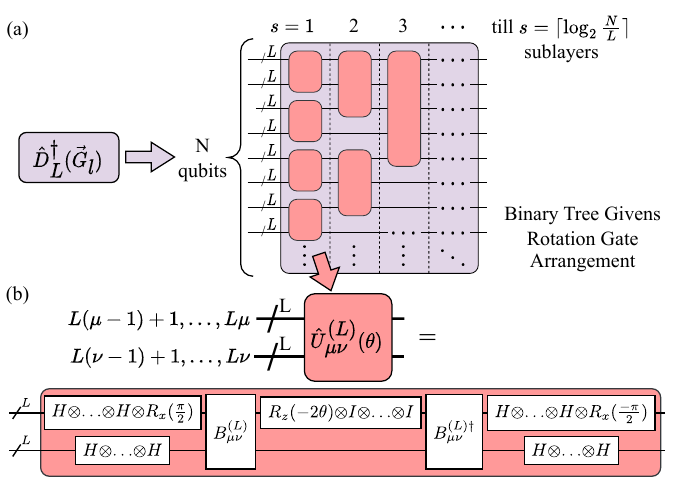}
\caption{(a) A product of multiple Givens rotations $\hat{D}^{\dagger}_{L}$ that is composed of Givens rotation gates $\hat{U}^{(L)}_{\mu,\nu}$ in a binary tree arrangement on the quantum circuit. (b) $\hat{U}^{(L)}_{\mu,\nu}$ is decomposed using Pauli rotation gates $R_x$ that acts on qubit $L\mu$, $R_z$ that acts on qubit $L(\mu{-}1){+}1$, Hadamards $H$ and CNOT ladders $B^{(L)}_{\mu\nu}$ that act on all $2L{-}1$ qubits in $\{L(\mu{-}1){+}1,{\ldots},L\mu{-}1\}$ and $\{L(\nu{-}1){+}1,{\ldots},L\nu\}$, with $B^{(L)}_{\mu\nu}$ also acting $\nu-\mu$ additional qubits in $\{L\mu, L(\mu{+}1),{\ldots}(\textrm{mod } L){\ldots},L(\nu{-}1)\}$. $\theta$ and $\vec{G}$ are scalar and vector parameters respectively.}
\label{fig3}
\end{figure}

\section{Extension to Correlated Ansatz}
Next, we extended the technique above by introducing a method to incorporate $L$-wise correlations into the fermionic ansatz state preparation, where $L{=}1$ reduces to the Slater determinant case. The idea is to use a set of anti-commuting operators that contains non-particle preserving multi-body Fock operators for the Clifford loaders in Eq.~\eqref{eq:cliff_load_def} and apply the same technique undertaken in Slater determinant case. Here for simplicity we work with Pauli string operators under the Jordan-Wigner mapping. We extend the anti-commuting operator $\hat{p}_\mu{=}\bigotimes_{r=1}^{\mu-1}\hat{Z}_{r}\hat{X}_\mu$ used to prepare the Slater determinant previously to
\be
\hat{p}^{(L)}_\mu=\bigotimes_{r=L(\textrm{mod }L)}^{L(\mu-1)}\hat{Z}_{r} \bigotimes_{r^\prime=L(\mu-1)+1}^{\mu L}\hat{X}_{r^\prime},
\ee
\noindent
to generate $L$-wise correlations, where $r$ index here increments with size $L$ from $L$ to $L(\mu{-}1)$. The modified operator $\hat{p}^{(L)}_\mu$ has $L$ Pauli-X terms and $\mu$ Pauli-Z terms with modulo $L$ indexes such that $\hat{p}^{(L)}_\mu$ remains anti-commuting $\{\hat{p}^{(L)}_\mu,\hat{p}^{(L)}_\nu\}{=}2\delta_{\mu\nu}\mathbf{I}$ as shown in Appendix~\ref{apx:anti_comm}. For example, the $L{=}2$ pairwise correlated anti-commuting operators is given as $\hat{p}^{(2)}_\mu=\hat{Z}_{2}\hat{Z}_{4}\hat{Z}_{6}\ldots\hat{Z}_{2\mu-2}\hat{X}_{2\mu-1}\hat{X}_{2\mu}$. Thus, we may prepare an $N$ mode $d$-occupied $L$-wise correlated state $\left|\Psi_{L}\right\rangle$ by applying $\frac{d}{L}$ Clifford loaders $\hat{C}_L$ 
\be
\left|\Psi_{L}(G)\right\rangle = \prod_{l=1}^{\frac{d}{L}} \hat{C}_L(\vec{G}_l)|\textrm{vac}\rangle, \label{eq:correlated_ansatz_clifford}
\ee
\noindent
where $G$ is a $\frac{N}{L}{\times}\frac{d}{L}$ orthonormal matrix, $\vec{G}_l$ is a column vector of $G$, which simplifies to
\be
\left|\Psi_{L}(G)\right\rangle \simeq \sum_{|B^\prime|=\frac{d}{L} } \textrm{det}(G_{B^\prime}) |B^\prime_L\rangle, \label{eq:corr_subspace_form}
\ee
\noindent
up to an unobserved global phase, where the sum is over all combinations of the ordered set $B^\prime$ containing $\frac{d}{L}$ unique integers between $1$ and $\frac{N}{L}$, $G_{B^\prime}$ is a $\frac{d}{L}{\times}\frac{d}{L}$ matrix minor of $G$ whose rows are restricted to $B^\prime$, $B^\prime_L{=}\{L(j{-}1){+}1, L(j{-}1){+}2,{\ldots}, Lj| j{\in} B^\prime \}$, and $|B^\prime_L\rangle$ denotes an $N$-mode Fock state with occupied modes indexed by $B^\prime_L$ as shown in Appendix~\ref{apx:cliff_load}.

The $L$-wise correlated ansatz state $\left|\Psi_{L}\right\rangle$ in Eq.~\eqref{eq:corr_subspace_form} is similar to the Slater determinant $\left|\Psi_{1}\right\rangle$ from Eq.~\eqref{eq:slater_subspace} in how the amplitudes are calculated, but differs in the Fock states that have non-zero amplitudes. In the Slater determinant case, all Fock states with $d$ particle number will have non-zero amplitudes, while in the $L{=}2$ pairwise case, all Fock states that have both $d$ particle number and $L{=}2$-tuple of neighboring occupations and neighboring vacants will have non-zero amplitudes. For instance in the case $N{=}4$ and $d{=}2$, Fock states $\{|0011\rangle, |1100\rangle\}$ will have non-zero amplitudes, and $\{|0101\rangle, |0110\rangle, |1010\rangle, |1001\rangle\}$ will have zero amplitudes. 

For a given normalized column $\vec{G}_l$, we may define a corresponding $L$-wise correlated Clifford loader $\hat{C}_{L}(\vec{G}_l) =\hat{D}_{L}^\dagger(\vec{G}_l)\hat{p}^{(L)}_1\hat{D}_{L}(\vec{G}_l)$ where $\hat{p}^{(L)}_1{=}\bigotimes_{r=1}^{L}\hat{X}_{r}$ consists of $L$ Pauli X gates that acts on the first $L$ qubits sandwiched two products of multiple Givens rotations $\hat{D}_{L}$, $\hat{D}^\dagger_{L}$. Each $\hat{D}_{L}$ is composed of Givens rotations gates $\hat{U}^{(L)}_{\mu\nu}(\theta){=}\exp[\theta\hat{p}^{(L)}_\mu\hat{p}^{(L)}_\nu]$ arranged in a similar binary tree pattern as shown in Fig.~\ref{fig3}(a). Since the Givens rotation gate $\hat{U}^{(L)}_{\mu\nu}(\theta)$ is a Pauli-string rotation gate, its gate decomposition~\cite{whitfieldSimulationElectronicStructure2011} and the corresponding rotation angle can be obtained in a similar fashion as in the Slater determinant case as shown Fig.~\ref{fig3}(b). We refer readers to Appendix~\ref{apx:givens_decomp} for the explicit form of the Givens Rotation in terms of Pauli operators.

\begin{figure}
\centering
\includegraphics[width=1\columnwidth, page=1]{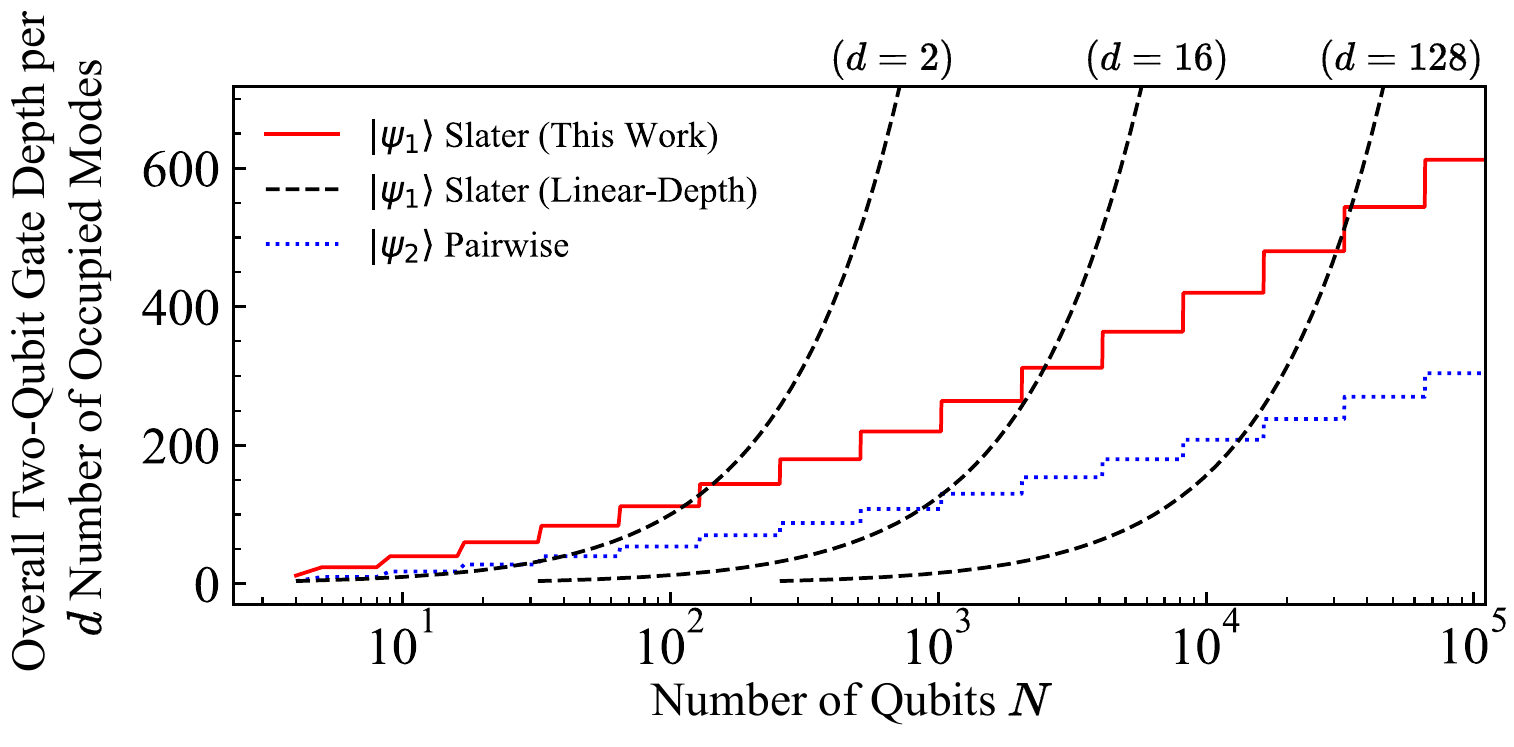}
\caption{Estimated two-qubit gate depth per occupied mode $d$ to prepare a $N$ mode Slater determinant $|\Psi_{1}\rangle$ and a $L{=}2$ pairwise correlated ansatz state $|\Psi_{2}\rangle$ using Clifford loaders compared to existing $d$-independent linear-depth approaches.}
\label{fig4}
\end{figure}

\section{Resource Analysis}
For simplicity, we treat all types of two-qubit gates depths as equal and assume no circuit compilation. We estimated that the overall two-qubit gate depth required to prepare a $L$-wise correlated ansatz state $|\Psi_{L}\rangle$ is $\frac{2d}{L}\left(\left\lceil\log^2_2 \frac{N}{L}\right\rceil{+}(1{+}2\log_2 L)\left\lceil\log_2 \frac{N}{L}\right\rceil \right){\approx}\mathcal{O}(\frac{d}{L}\log_2^2 \frac{N}{L})$ as shown in Appendix~\ref{apx:scaling}. This shows that our approach to prepare a $L$-wise correlated ansatz state $\left|\Psi_{L}\right\rangle$ is shallower than a Slater determinant $\left|\Psi_{1}\right\rangle$ by at least a factor of $L$, \textit{ceteris paribus}.

We plotted the estimated two-qubit gate depth per occupied mode for Slater determinants $|\Psi_{1}\rangle$ and pairwise correlated ansatz states $|\Psi_{2}\rangle$ on a quantum computer in Fig.~\ref{fig4}, and compared against the previous linear depth approach of preparing $d{=}2,16$ and $128$-occupied Slater determinants, which has a $d$-independent two-qubit gate depth of about $2N$~\cite{weckerSolvingStronglyCorrelated2015, kivlichanQuantumSimulationElectronic2018, aruteHartreeFockSuperconductingQubit2020}. Indeed, the crossover point $d{\approx}\frac{N}{\log^2_2 N}$ shows that the minimum number of qubits $N$ required to achieve a shallower circuit increases sub-exponentially with the number of occupied modes $d$. However, we highlight that this crossover point can be practically surpassed by near-term quantum devices such as superconducting qubits~\cite{kjaergaardSuperconductingQubitsCurrent2020} and trapped-ions~\cite{bruzewiczTrappedionQuantumComputing2019} with less than $10^{5}$ qubits for systems with $2{\le}d{\le}128$ occupied fermionic modes, which is a sizable range that encompass many systems of interest in quantum chemistry and condensed matter physics. In general, our approach is suitable for problems classes that have a sufficiently slow growing $d{\ll}{\mathcal{O}}{\left(\frac{N}{\log^2_2 N}\right)}$. One such problem is computation of quantum observable quantities for fermionic systems for the complete basis set limit, where $d$ is preserved, but said quantities are computed for increasing values of $N$ and extrapolated using various schemes to very large limits of $N$~\cite{martinInitioTotalAtomization1996, halkierBasissetConvergenceEnergy1999, spackmanEstimatingCCSDBasisset2015}.

\begin{figure}
\centering
\includegraphics[width=1\columnwidth, page=1]{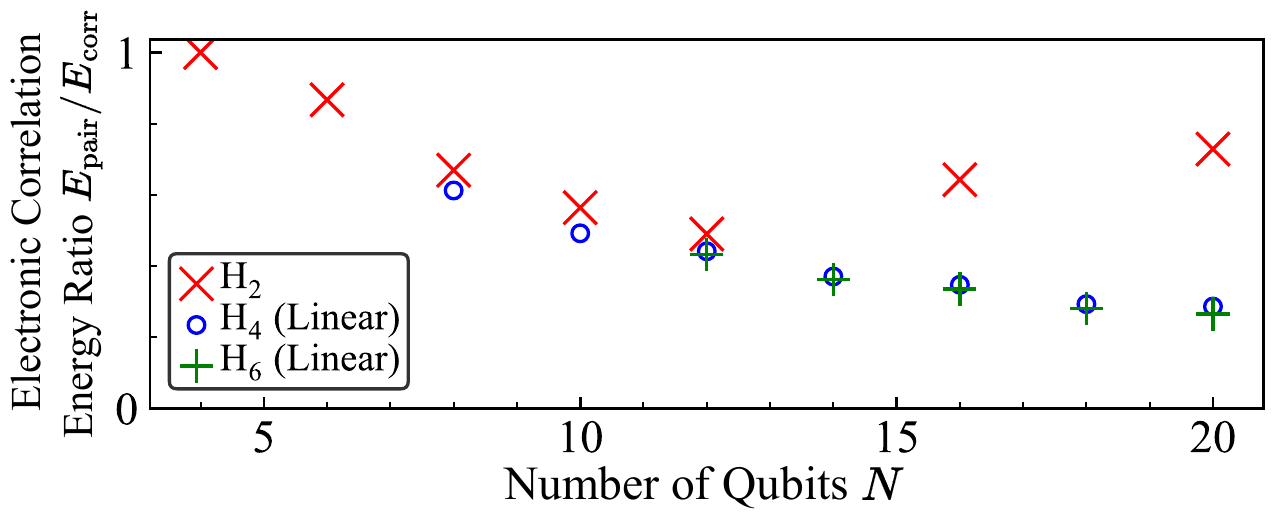}
\caption{Numerically calculated fraction of the electronic correlation energy ratio $E_{\textrm{pair}}/E_{\textrm{corr}}$ captured by an optimized pairwise correlated ansatz state for hydrogen chains up to $\textrm{H}_6$ at fixed interatomic distance of 1.4 bohr for various basis set mixtures up to 20 qubits.}
\label{fig5}
\end{figure}

% Result: Correlation Energy
\section{Example: Linear Hydrogen Molecular Chains}
To validate our $L{=}2$ pairwise correlated ansatz state, we numerically evaluated the fraction of the electronic correlation energy $E_{\textrm{pair}}/E_{\textrm{corr}}$ captured by the optimized pairwise correlated ansatz state $\left|\Psi_{2}(G^*)\right\rangle$ for three linear hydrogen molecular chains (H$_2$, H$_4$ and H$_6$) with ($d{=}2$, $4$, $6$) electrons at a fixed interatomic distance of 1.4 bohr, where pairwise electronic correlation is likely to be significant. $E_{\textrm{pair}}{=}E_{\textrm{HF}}{-}\langle\hat{H}_e\rangle$ is the correlation energy beyond the mean-field energy $E_{\textrm{HF}}$ of a molecule captured by $\left|\Psi_{2}(G^*)\right\rangle$, while $E_{\textrm{corr}}{=}E_{\textrm{HF}}{-}E_{\textrm{FCI}}$ is the exact value of the correlation energy, where $E_{\textrm{FCI}}$ is known as the full configuration interaction (FCI) energy. $G^*$ is an optimized parameter matrix, obtained using a classical quasi-Newton L-BFGS-B optimizer that minimizes the expectation of the electronic Hamiltonian $\langle\hat{H}_e\rangle$. We considered different mixtures of atomic basis sets (STO-3G, 6-31G, 6-311G, cc-pvdz, aug-cc-pvdz) for each hydrogen atom, resulting in system sizes ranging from 4 to 20 qubits. Such mixing of basis sets for each atom is a common strategy in computational quantum chemistry to reduce the resources required to achieve a desired precision~\cite{plascenciaImportanceLigandBasis2017}. All calculations were performed numerically using SciPy~\cite{virtanenSciPyFundamentalAlgorithms2020}, PYSCF~\cite{sunPySCFPythonbasedSimulations2018} and Pennylane~\cite{bergholmPennyLaneAutomaticDifferentiation2022}. Fig.~\ref{fig5} shows that a large fraction of the electronic correlation energy $E_{\textrm{pair}}/E_{\textrm{corr}}$ is captured by the optimized pairwise correlated ansatz state $\left|\Psi_{2}(G^*)\right\rangle$.

\section{Summary and outlook}
We have proposed Givens rotation-based Clifford loaders for efficient preparation of $d$-occupied Slater determinants $\left|\Psi_{1}\right\rangle$ of $N$ modes using shallower $\mathcal{O}(d\log^2_2 N)$ two-qubit gate depth quantum circuits. We have also showed that by redefining new sets of the anti-commutation operators $\hat{p}_\mu$ for the Clifford loaders, the same technique can prepare $L$-wise correlated ansatz states $\left|\Psi_{L}\right\rangle$ to yield shallower quantum circuits than that of Slater determinants by at least a factor of $L$. As demonstrated in the application of $L{=}2$ pairwise correlated ansatz states to hydrogen chains, $L$-wise correlated states are potentially useful in fermionic systems with significant $L$-wise fermionic correlation, even though it is not expected to fully capture all the correlation energy. It will be interesting to generalize the Clifford loaders to other types of fermionic correlation while keeping the same shallow gate depth scaling intact.

To the best of our knowledge, our approach to fermionic ansatz state preparation offers an subexponential improvement in gate depth over existing methods in the second quantization with respect to system size $N$, for fermionic problems where the number of occupied modes is $d{\ll}\mathcal{O}{\left(\frac{N}{\log^2_2 N}\right)}$. Nonetheless, our results have established Clifford loaders via Givens rotations as an efficient, yet practical and scalable fermionic ansatz state preparation technique, which will enable the study of molecules and materials requiring larger basis set sizes on near-term quantum devices.

\begin{acknowledgments}
This research is supported by the National Research Foundation, Singapore and A*STAR under its CQT Bridging Grant and Quantum Engineering Programme NRF2021-QEP2-02-P02, A*STAR (\#21709) and by EU HORIZON - Project 101080085 — QCFD.
\end{acknowledgments}

% \clearpage
% \onecolumn
\onecolumngrid %revtex only
\newpage

\appendix
\setcounter{figure}{0}                       
\renewcommand\thefigure{A.\arabic{figure}}   

\section{Proof of Equivalence between Two Definitions of an Arbitrary Slater Determinant}
\label{apx:equivalence_slater}

An arbitrary Slater determinant $|\Psi_{1}(A)\rangle$ with $d$ occupied and $N-d$ unoccupied fermionic modes is defined as
\be
|\Psi_{1}(A)\rangle = \prod_{l=1}^{d}\sum^N_{k=1} A_{lk} \hat{a}^\dagger_k  |\textrm{vac}\rangle, \label{eq:apx_slater_def}
\ee

where $A$ is an $N{\times}d$ real matrix such that all the $d$ columns are orthogonal and normalized, $|\textrm{vac}\rangle$ is a vacuum state, and $\hat{a}^\dagger_k$ is a creation operator acting on the $k$\textsuperscript{th} mode. 
To begin, we expand the product in Eq.~\eqref{eq:apx_slater_def} and replace the index $k$ in the summation with indices $k_1,\ldots,k_d$,
\begin{align}
|\Psi_{1}(A)\rangle  &= \left( \sum^N_{k_d=1} A_{dk_d} \hat{a}^\dagger_{k_d} \right) \ldots \left( \sum^N_{k_1=1} A_{1k_1} \hat{a}^\dagger_{k_1} \right) |\textrm{vac}\rangle \\
 &= \sum^N_{k_d,\ldots,k_1=1} A_{dk_d} \ldots A_{1k_1} \hat{a}^\dagger_{k_d} \ldots \hat{a}^\dagger_{k_1}  |\textrm{vac}\rangle, \label{eq:apx_slater_final}
\end{align}

We want to show that Eq.~\eqref{eq:apx_slater_final} is equivalent to the alternate definition of the Slater determinant,
\be
\left|\Psi_{\textrm{1,alt}}(A)\right\rangle = \prod_{l=1}^{d}\sum_{k=1}^{N}A_{lk}\left(\hat{a}^\dagger_k+\hat{a}_k\right) |\textrm{vac}\rangle, \label{eq:apx_slater_alt_def}
\ee
where we use an anti-commuting operator $\hat{a}^\dagger_k{+}\hat{a}_k$. Starting from Eq.~\eqref{eq:apx_slater_alt_def}, we expand the product and replace index $k$ in the summation with indices $k_1,\ldots,k_d$,
\be
\left|\Psi_{\textrm{1,alt}}(A)\right\rangle = \left(\sum_{k_d=1}^{N}A_{dk_d}\left(\hat{a}^\dagger_{k_d}+\hat{a}_{k_d}\right)\right)\ldots \left(\sum_{k_1=1}^{N}A_{1k_1}\left(\hat{a}^\dagger_{k_1}+\hat{a}_{k_1}\right)\right)|\textrm{vac}\rangle \label{eq:apx_slater_alt_expand}
\ee

Here, we consider evaluating the product of the rightmost two summation terms in Eq.~\eqref{eq:apx_slater_alt_expand}. We split the derivation into two cases $d{=}1$ and $d{>}1$. For $d{=}1$, we note that $\hat{a}_{j}|\textrm{vac}\rangle{=}0$ for any mode $j$, thus both Eq.~\eqref{eq:apx_slater_final} and~\eqref{eq:apx_slater_alt_expand} become trivially equivalent. For $d{>}1$ we have,
\begin{align}
&\left(\sum_{k_2=1}^{N}A_{2k_2}\left(\hat{a}^\dagger_{k_2}+\hat{a}_{k_2}\right)\right) 
\left(\sum_{k_1=1}^{N}A_{1k_1}\left(\hat{a}^\dagger_{k_1}+\hat{a}_{k_1}\right)\right) \nonumber \\
&= \sum_{k_2,k_1=1}^{N}A_{2k_2}A_{1k_1}\left(\hat{a}^\dagger_{k_2}+\hat{a}_{k_2}\right)\left(\hat{a}^\dagger_{k_1}+\hat{a}_{k_1}\right)\\
& = \sum_{k_2,k_1=1}^{N}A_{2k_2}A_{1k_1}\left(\hat{a}^\dagger_{k_2}\hat{a}^\dagger_{k_1}+\hat{a}^\dagger_{k_2}\hat{a}_{k_1}+\hat{a}_{k_2}\hat{a}^\dagger_{k_1}+\hat{a}_{k_2}\hat{a}_{k_1}\right), \label{eq:apx_alt_expand_1}
\end{align}
we then apply fermionic commutation relation $\{\hat{a}_\alpha,\hat{a}^\dagger_\beta\}{=}\delta_{\alpha\beta}\mathbf{I}$ to Eq.~\eqref{eq:apx_alt_expand_1} to get,
\be
=\sum_{k_2,k_1=1}^{N}A_{2k_2}A_{1k_1}\left(\hat{a}^\dagger_{k_2}\hat{a}^\dagger_{k_1}+\delta_{k_2k_1}\mathbf{I}+\hat{a}_{k_2}\hat{a}_{k_1}\right),
\ee
and since the columns of $A$ are orthogonal, the inner product between any column $i,j$ vanishes $\sum_{k}A_{ik}A_{jk}=0$, we thus have
\be
=\sum_{k_2,k_1=1}^{N}A_{2k_2}A_{1k_1}\left(\hat{a}^\dagger_{k_2}\hat{a}^\dagger_{k_1}+\hat{a}_{k_2}\hat{a}_{k_1}\right). \label{eq:apx_product_result}
\ee

Substituting Eq.~\eqref{eq:apx_product_result} back into Eq.~\eqref{eq:apx_slater_alt_expand}, and using $\hat{a}_{j}|\textrm{vac}\rangle=0$ for any mode $j$ gives
\be
\left|\Psi_{\textrm{1,alt}}(A)\right\rangle = \left(\sum_{k_d=1}^{N}A_{dk_d}\left(\hat{a}^\dagger_{k_d}+\hat{a}_{k_d}\right)\right)\ldots \left(\sum_{k_2,k_1=1}^{N}A_{2k_2}A_{1k_1}\hat{a}^\dagger_{k_2}\hat{a}^\dagger_{k_1}\right)|\textrm{vac}\rangle. \label{eq:apx_slater_alt_expand1}
\ee
Henceforth we consider the even and odd $d$ cases separately. First, assuming $d$ is even we can reapply the result Eq.~\eqref{eq:apx_product_result} to the rest of the pairs of summation terms in Eq.~\eqref{eq:apx_slater_alt_expand1}, which yields
\begin{align}
\left|\Psi_{\textrm{1,alt}}(A)\right\rangle &= \left(\sum_{k_d,k_{d-1}=1}^{N}A_{dk_d}A_{d-1k_{d-1}}\hat{a}^\dagger_{k_d}\hat{a}^\dagger_{k_{d-1}}\right)\ldots \left(\sum_{k_2,k_1=1}^{N}A_{2k_1}A_{1k_1}\hat{a}^\dagger_{k_2}\hat{a}^\dagger_{k_1}\right)|\textrm{vac}\rangle \\
& = \sum_{k_d,\ldots,k_1=1}^{N}A_{dk_d}\ldots A_{1k_1}\hat{a}^\dagger_{k_d}\ldots \hat{a}^\dagger_{k_1} |\textrm{vac}\rangle. \label{eq:apx_slater_alt_expand_even_final}
\end{align}
Alternatively, if $d$ is odd we have
\begin{align}
& \left|\Psi_{\textrm{1,alt}}(A)\right\rangle \nonumber \\
&= \left(\sum_{k_d=1}^{N}A_{dk_d}\left(\hat{a}^\dagger_{k_d}+\hat{a}_{k_d}\right)\right)\sum_{k_{d-1},\ldots,k_1=1}^{N}A_{d-1k_{d-1}}\ldots A_{1k_1}\hat{a}^\dagger_{k_{d-1}}\ldots \hat{a}^\dagger_{k_1} |\textrm{vac}\rangle \\
&=\sum_{k_d,\ldots,k_1=1}^{N}A_{dk_d}\ldots A_{1k_1}\left(\hat{a}^\dagger_{k_d}\hat{a}^\dagger_{k_{d-1}}\ldots \hat{a}^\dagger_{k_1} + \hat{a}_{k_d}\hat{a}^\dagger_{k_{d-1}}\ldots \hat{a}^\dagger_{k_1} \right)|\textrm{vac}\rangle \label{eq:apx_slater_alt_expand_odd1} \\
&=\sum_{k_d,\ldots,k_1=1}^{N}A_{dk_d}\ldots A_{1k_1}\left(\hat{a}^\dagger_{k_d}\hat{a}^\dagger_{k_{d-1}}\ldots \hat{a}^\dagger_{k_1} +\cancelto{0}{\delta_{k_dk_{d-1}}\hat{a}^\dagger_{k_{d-2}}\ldots \hat{a}^\dagger_{k_1}} - \hat{a}^\dagger_{k_{d-1}}\hat{a}_{k_d}\hat{a}^\dagger_{k_{d-2}}\ldots \hat{a}^\dagger_{k_1} \right)|\textrm{vac}\rangle \label{eq:apx_slater_alt_expand_odd2} \\
&=\sum_{k_d,\ldots,k_1=1}^{N}A_{dk_d}\ldots A_{1k_1}\left(\hat{a}^\dagger_{k_d}\hat{a}^\dagger_{k_{d-1}}\ldots \hat{a}^\dagger_{k_1} -\cancelto{0}{\hat{a}^\dagger_{k_{d-1}}\delta_{k_{d}k_{d-2}}\hat{a}^\dagger_{k_{d-3}}\ldots \hat{a}^\dagger_{k_1}} + \cancelto{0}{\hat{a}^\dagger_{k_{d-1}}\hat{a}^\dagger_{k_{d-2}}\hat{a}_{k_d}\hat{a}^\dagger_{k_{d-3}}\ldots \hat{a}^\dagger_{k_1}}\quad\, \right)|\textrm{vac}\rangle \label{eq:apx_slater_alt_expand_odd3} \\
& = \sum_{k_d,\ldots,k_1=1}^{N}A_{dk_d}\ldots A_{1k_1}\hat{a}^\dagger_{k_d}\ldots \hat{a}^\dagger_{k_1} |\textrm{vac}\rangle, \label{eq:apx_slater_alt_expand_odd_final}
\end{align}
where we have applied fermionic commutation relation in Eq.~\eqref{eq:apx_slater_alt_expand_odd1} and matrix orthogonality in Eq.~\eqref{eq:apx_slater_alt_expand_odd2},~\eqref{eq:apx_slater_alt_expand_odd3}. Hence, by combining the even Eq.~\eqref{eq:apx_slater_alt_expand_even_final} and odd Eq.~\eqref{eq:apx_slater_alt_expand_odd_final} results, we have established the equivalence between the alternative definition in Eq.~\eqref{eq:apx_slater_alt_def} to the original definition in Eq.~\eqref{eq:apx_slater_def}.

% \newpage
\section{Proof of Anti-Commutation Relations}
\label{apx:anti_comm}
Here, we shall show that $\hat{p}_k = \hat{a}^\dagger_k+\hat{a}_k$ has the desired anti-commutation relation $\{\hat{p}_i,\hat{p}_j\}{=}2\delta_{ij}\mathbf{I}$. 
\begin{align}
\{\hat{p}_i,\hat{p}_j\} &= \{\hat{a}^\dagger_i+\hat{a}_i,\hat{a}^\dagger_j+\hat{a}_j\}\\\
&=(\hat{a}^\dagger_i+\hat{a}_i)(\hat{a}^\dagger_j+\hat{a}_j)+(\hat{a}^\dagger_j+\hat{a}_j)(\hat{a}^\dagger_i+\hat{a}_i)\\
&= \cancel{\hat{a}^\dagger_i\hat{a}^\dagger_j}+\hat{a}_i\hat{a}^\dagger_j+\hat{a}^\dagger_i\hat{a}_j+\cancel{\hat{a}_i\hat{a}_j}+\cancel{\hat{a}^\dagger_{j}\hat{a}^\dagger_i}+\hat{a}_{j}\hat{a}^\dagger_i+\hat{a}^\dagger_{j}\hat{a}_i+\cancel{\hat{a}_{j}\hat{a}_i}\\
&= \hat{a}_i\hat{a}^\dagger_j+\hat{a}^\dagger_i\hat{a}_j+\hat{a}_{j}\hat{a}^\dagger_i+\hat{a}^\dagger_{j}\hat{a}_i\\
 &= 2\delta_{ij}I \quad \textrm{(shown)}
\end{align}

Next, we shall show that $\hat{p}^{(L)}_\mu{=}\bigotimes_{r=L(\textrm{mod }L)}^{L(\mu-1)}\hat{Z}_{r} \bigotimes_{r^\prime=L(\mu-1)+1}^{\mu L}\hat{X}_{r^\prime}$, used to incorporate $L$-wise correlation into the Clifford loaders via the Givens rotation approach has the desired anti-commutation relation $\{\hat{p}^{(L)}_\mu,\hat{p}^{(L)}_\nu\}=2\delta_{\mu\nu}\mathbf{I}$. We shall split the derivation into two cases $\mu{=}\nu$ and $\mu{<}\nu$. Let us first consider $\mu{=}\nu$, where
\begin{align}
2\hat{p}_\mu^{(L)}\hat{p}_\mu^{(L)} &=2\left(\bigotimes_{r=L(\textrm{mod }L)}^{L(\mu-1)}\hat{Z}_{r} \bigotimes_{r^\prime=L(\mu-1)+1}^{\mu L}\hat{X}_{r^\prime}\right) \left(\bigotimes_{s=L(\textrm{mod }L)}^{L(\mu-1)}\hat{Z}_{s} \bigotimes_{s^\prime=L(\mu-1)+1}^{\mu L}\hat{X}_{s^\prime}\right) \\
&=2\mathbf{I}\,\textrm{(shown)}
\end{align}
Second, without loss of generality, let us consider $\mu{<}\nu$ and we note that $\hat{X}_\mu\hat{Z}_\mu=-i\hat{Y}_\mu$,
\begin{align}
\hat{p}_\mu^{(L)}\hat{p}_\nu^{(L)} &=\left(\bigotimes_{r=L(\textrm{mod }L)}^{L(\mu-1)}\hat{Z}_{r} \bigotimes_{r^\prime=L(\mu-1)+1}^{\mu L}\hat{X}_{r^\prime}\right) \left(\bigotimes_{s=L(\textrm{mod }L)}^{L(\nu-1)}\hat{Z}_{s} \bigotimes_{s^\prime=L(\nu-1)+1}^{\nu L}\hat{X}_{s^\prime}\right)\\
&=-i\left[\bigotimes_{r=(\mu-1)L+1}^{\mu L-1}\hat{X}_{r} \hat{Y}_{\mu L}\bigotimes_{r^\prime=(\mu+1)L(\textrm{mod }L)}^{(\nu-1)L}\hat{Z}_{r^\prime}\bigotimes_{r^{\prime\prime}=(\nu-1)L+1}^{\nu L}\hat{X}_{r^{\prime\prime}}\right]
\end{align}
Therefore, by noting $\hat{Z}_\mu\hat{X}_\mu=i\hat{Y}_\mu$ we then have,
\begin{align}
\{\hat{p}_\mu^{(L)},\hat{p}_\nu^{(L)}\} &=\hat{p}_\mu^{(L)}\hat{p}_\nu^{(L)}+\hat{p}_\nu^{(L)}\hat{p}_\mu^{(L)}\\
&=-i\left[\bigotimes_{r=(\mu-1)L+1}^{\mu L-1}\hat{X}_{r} \hat{Y}_{\mu L}\bigotimes_{r^\prime=(\mu+1)L(\textrm{mod }L)}^{(\nu-1)L}\hat{Z}_{r^\prime}\bigotimes_{r^{\prime\prime}=(\nu-1)L+1}^{\nu L}\hat{X}_{r^{\prime\prime}}\right]\nonumber\\&\quad+i\left[\bigotimes_{s=(\mu-1)L+1}^{\mu L-1}\hat{X}_{s} \hat{Y}_{\mu L}\bigotimes_{s^\prime=(\mu+1)L(\textrm{mod }L)}^{(\nu-1)L}\hat{Z}_{s^\prime}\bigotimes_{s^{\prime\prime}=(\nu-1)L+1}^{\nu L}\hat{X}_{s^{\prime\prime}}\right]\\
&= 0 \quad \textrm{(shown)}
\end{align}

% \newpage
\section{Evaluation of the Slater Determinant and \texorpdfstring{$L$}{L}-wise Correlated Ansatz using Clifford Loaders}
\label{apx:cliff_load}
In the main text, we proposed to construct Clifford loaders $\hat{C}$ using Givens rotation to prepare Slater determinant $|\Psi_{1}(A)\rangle$ as follows
\be
|\Psi_{1}(A)\rangle = \prod_{l=1}^{d} \hat{C}(\vec{A}_l)  |\textrm{vac}\rangle. \label{eq:qpx_cliff_loader_slater}
\ee

We want to show that Eq.~\eqref{eq:qpx_cliff_loader_slater} can be mathematically evaluated using geometric algebra, also known as real Clifford algebra. We define a Clifford loader $\hat{C}(\vec{x})$ for a given normalized sized-$N$ vectors $\vec{x}$ as a linear combination of anti-commuting operators $\hat{p}_r$ as follows
\be
\hat{C}(\vec{x}) = \sum_{\mu=1}^N x_\mu \hat{p}_\mu. \label{eq:apx_cliff_def}
\ee

A geometric product of two Clifford loaders $\hat{C}(\vec{x})\hat{C}(\vec{y})$ for any two normalized sized-$N$ vectors $\vec{x}$, $\vec{y}$ is defined as
\be
\hat{C}(\vec{x})\hat{C}(\vec{y})=\hat{C}(\vec{x}) \cdot \hat{C}(\vec{y}) + \hat{C}(\vec{x}) \wedge \hat{C}(\vec{y}). \label{eq:apx_geo_prod_def}
\ee
where $\cdot$ and $\wedge$ refers to the standard inner dot and exterior wedge product respectively. Substituting the definition Eq.~\eqref{eq:apx_cliff_def} into Eq.~\eqref{eq:apx_geo_prod_def}, we have
\be
\hat{C}(\vec{x})\hat{C}(\vec{y})=\sum_{r=1}^N x_r y_r  (\hat{p}_r \cdot \hat{p}_r) + \sum_{\mu,\nu=1}^N x_\mu y_\nu (\hat{p}_\mu \wedge \hat{p}_\nu). \label{eq:apx_geo_prod}
\ee

In this work, we shall only consider orthogonal and normalized sized-$N$ vectors $\vec{A}_l$ for $l=1,2,\ldots,d$, such that the inner product of any two vectors is zero. As a result, the geometric product of two Clifford loader is simply equivalent to the its exterior product as the first term of Eq.~\eqref{eq:apx_geo_prod} vanishes under orthogonality. Thus, products of multiple Clifford loaders can be easily written as  exterior product of mutiple anti-commuting operators
\be
\prod_{l=1}^{d} \hat{C}(\vec{A}_l) = \sum_{\mu,\nu,\ldots,r=1}^N A_{\mu1} A_{\nu2} \dots A_{rd} \underset{d \textrm{ operators}}{\underbrace{(\hat{p}_\mu \wedge \hat{p}_\nu \wedge \ldots \wedge \hat{p}_r )}}.  \label{eq:apx_ext_prod_anti_comm}
\ee
We note the following identities of exterior product
\be
\hat{p}_\mu \wedge \hat{p}_\mu = 0, \label{eq:apx_ext_iden1}
\ee
and 
\be
\hat{p}_{\sigma_1} \wedge \hat{p}_{\sigma_2} \wedge \ldots \wedge \hat{p}_{\sigma_d} =\textrm{sgn}(\sigma) \hat{p}_{B_1} \wedge \hat{p}_{B_2} \wedge \ldots \wedge \hat{p}_{B_d}, \label{eq:apx_ext_iden2}
\ee
where we let $\sigma$ be a permutation of $\{B_1,B_2,\ldots,B_d\}$ for any ordered set $B$ containing $d$ unique integers between $1$ and $N$, $B_\mu$ and $\sigma_\mu$ refers to the $\mu$\textsuperscript{th} integer of $B$ and  $\sigma$ respectively. Using the above identities~\eqref{eq:apx_ext_iden1} and~\eqref{eq:apx_ext_iden2}, the sum in Eq.~\eqref{eq:apx_ext_prod_anti_comm} reduces to 
\be
\prod_{l=1}^{d} \hat{C}(\vec{A}_l) = \sum_{|B|=d } \sum_{\sigma \in B } \textrm{sgn}(\sigma) A_{\sigma_1 1} A_{\sigma_2 2} \dots A_{\sigma_d d} \left( \hat{p}_{B_1} \wedge \hat{p}_{B_2} \wedge \ldots \wedge \hat{p}_{B_d}\right),
\ee
where the outer sum is over all possible combinations of the ordered set $B$ containing $d$-unique integers between $1$ and $N$ and the inner sum is over all possible integer permutation $\sigma$ of each $B$. Using the Leibniz determinant formula for matrix minors
\be
\textrm{det}(A_B) = \sum_{\sigma \in B } \textrm{sgn}(\sigma) A_{\sigma_1 1} A_{\sigma_2 2} \dots A_{\sigma_d d}, \label{eq:apx_det_def}
\ee
where $A_B$ is a $d{\times}d$ matrix minor of $A$ whose rows are restricted to $B$, then we have,
\be
\prod_{l=1}^{d} \hat{C}(\vec{A}_l) = \sum_{|B|=d } \textrm{det}(A_B)  \hat{p}_{B_1} \wedge \hat{p}_{B_2} \wedge \ldots \wedge \hat{p}_{B_d}.\label{eq:apx_cliff_wedge}
\ee

Hence, by letting the anti-commuting operator be
\be
\hat{p}_\mu=\hat{a}^\dagger_\mu+\hat{a}_\mu \label{eq:apx_anti_comm_ops_slater}
\ee
\noindent
and applying the product of $d$ Clifford loaders~\eqref{eq:apx_cliff_wedge} onto a vacuum state $|\textrm{vac}\rangle$, we obtain the alternative expression of the Slater determinant
\be
\prod_{l=1}^{d} \hat{C}(\vec{A}_l) |\textrm{vac}\rangle = \sum_{|B|=d } \textrm{det}(A_B) |B\rangle,
\ee
where $|B\rangle$ denotes a Fock state whose occupied modes are indexed by $B$. 

Next, we consider extending the application of Clifford loaders to prepare $L$-wise correlated ansatz states, where $L{=}1$ reduces to the Slater determinant case. The idea is to use a new set of anti-commuting operators $\hat{p}^{(L)}_\mu$ that contains non-particle preserving multi-body Fock operators for the Clifford loaders in Eq.~\eqref{eq:apx_cliff_def}. In the main text, under the Jordan-Wigner mapping, we modify the anti-commuting operator $\hat{p}_\mu{=}\bigotimes_{r=1}^{\mu-1}\hat{Z}_{r}\hat{X}_\mu$ in Eq.~\eqref{eq:apx_anti_comm_ops_slater} used to prepare the Slater determinant above to become
\be
\hat{p}^{(L)}_\mu=\bigotimes_{r=L(\textrm{mod }L)}^{L(\mu-1)}\hat{Z}_{r} \bigotimes_{r^\prime=L(\mu-1)+1}^{\mu L}\hat{X}_{r^\prime}, \label{eq:apx_anti_comm_ops_corr}
\ee
\noindent
to apply $L$-wise correlation, where $r$ index here increments with size $L$ from $L$ to $L(\mu{-}1)$. This modified operator $\hat{p}^{(L)}_\mu$ has $L$ Pauli-X terms and $\mu$ Pauli-Z terms with modulo $L$ indexes such that $\{\hat{p}^{(L)}_\mu,\hat{p}^{(L)}_\nu\}{=}2\delta_{\mu\nu}\mathbf{I}$. For example, the $L{=}2$ pairwise correlated anti-commuting operators is given as $\hat{p}^{(2)}_\mu=\hat{Z}_{2}\hat{Z}_{4}\hat{Z}_{6}\ldots\hat{Z}_{2\mu-2}\hat{X}_{2\mu-1}\hat{X}_{2\mu}$. In terms of Fermionic creation and annihilation operators, Eq.~\eqref{eq:apx_anti_comm_ops_corr} maps to back to
\be
\hat{p}^{(L)}_\mu=\begin{dcases}\bigotimes_{r=L(\textrm{mod }L)}^{L(\mu-1)}\left(\mathbf{I} - 2 \hat{a}^\dagger_r \hat{a}_r\right) \bigotimes_{r^\prime=L(\mu-1)+1}^{\mu L} \left(\hat{a}^\dagger_r+(-1)^{r^\prime}\hat{a}_r\right)&  \textrm{if } L \textrm{ is even},\\
-\bigotimes_{r=1\backslash(\textrm{mod }L)}^{L(\mu-1)-1}\left(\mathbf{I} - 2 \hat{a}^\dagger_r \hat{a}_r\right) \bigotimes_{r^\prime=L(\mu-1)+1}^{\mu L} \left(\hat{a}^\dagger_r+(-1)^{r^\prime-L(\mu-1)-1}\hat{a}_r\right)& \textrm{if } L \textrm{ is odd},
\end{dcases}\label{eq:apx_anti_comm_ops_corr_fock}
\ee
\noindent
where $r$ index in the odd $L$ case increments with size 1 from $1$ to $L(\mu{-}1){-}1$ but skips every index that is a multiple of $L$. Here we note in Eq.~\eqref{eq:apx_anti_comm_ops_corr_fock}, all fermionic terms when in normal ordered form that contain any annihilation operators will vanish when acted upon a vacuum state and fermionic terms that contain only creation operators will survive. Thus, by substituting Eq.~\eqref{eq:apx_anti_comm_ops_corr_fock} into the product of $\frac{d}{L}$ Clifford loaders Eq.~\eqref{eq:apx_cliff_wedge} and applying it onto a vacuum state $|\textrm{vac}\rangle$, we obtain an expression of the $L$-wise correlated ansatz state up to an unobserved global phase
\be
\prod_{l=1}^{\frac{d}{L}} \hat{C}_L(\vec{G}_l) |\textrm{vac}\rangle \simeq \sum_{|B^\prime|=\frac{d}{L} } \textrm{det}(G_{B^\prime}) |B^\prime_L\rangle,
\ee
where the sum is over all possible combinations ordered set $B^\prime$ containing $\frac{d}{L}$ unique integers between $1$ and $\frac{N}{L}$, $G_{B^\prime}$ is a $\frac{d}{L}{\times}\frac{d}{L}$ matrix minor of $G$ whose rows are restricted to $B^\prime$, $B^\prime_L=\{L(j-1)+1, L(j-1)+2,\ldots, Lj| j\in B^\prime \}$ and $|B^\prime_L\rangle$ denotes a $N$-mode Fock basis whose occupied mode are index by $B^\prime_L$.

% \newpage
\section{Givens Rotation Gate and its Decomposition for the \texorpdfstring{$L$}{L}-wise Correlated Ansatz State}
\label{apx:givens_decomp}
\begin{figure}[htp]
\centering
\includegraphics[width=0.6\columnwidth, page=1]{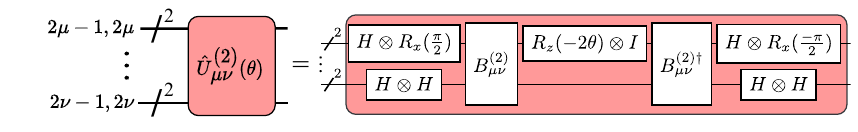}
\caption{The pairwise Givens rotation gate $\hat{U}^{(2)}_{\mu,\nu}$ is decomposed using Pauli rotation gates $R_x$ that acts on qubit $2\mu$ that correspond to $\hat{Y}_{2\mu}$, $R_z$ that acts on qubit $2\mu{-}1$, Hadamards $H$ and CNOT ladders $B^{(L)}_{\mu\nu}$ that acts on all $3$ qubits in $\{2\mu{-}1,2\nu{-}1,2\nu\}$ that corresponds to correspond to $\hat{X}_{2\mu-1}\hat{X}_{2\nu-1}\hat{X}_{2\nu}$, with $B^{(L)}_{\mu\nu}$ also acting $\nu-\mu$ additional qubits in $\{2\mu, 2(\mu{+}1),{\ldots}(\textrm{mod } 2){\ldots},2(\nu{-}1)\}$. $\theta$ is a scalar parameter.}
\label{apx_fig1}
\end{figure}

Given rotation gate is defined in the main text as 
\be
\hat{U}^{(L)}_{\mu\nu}(\theta)=\exp[\theta\hat{p}^{(L)}_\mu\hat{p}^{(L)}_\nu]
\ee
where using anti-commuting operators in Eq.~\eqref{eq:apx_anti_comm_ops_corr}, it becomes  
\be
\hat{U}^{(L)}_{\mu\nu}(\theta) = \exp\left[-i\theta\bigotimes_{r=L(\mu{-}1){+}1}^{\mu L{-}1}\hat{X}_{r} \hat{Y}_{\mu L} \bigotimes_{r^{\prime}=L(\nu-1)+1}^{\nu L}\hat{X}_{r^{\prime}} \bigotimes_{\substack{r^{\prime\prime}=L(\mu+1)\\(\textrm{mod }L)}}^{L(\nu{-}1)}\hat{Z}_{r^{\prime\prime}}\right]. \label{eq:generator}
\ee
\noindent
The $r^{\prime\prime}$ index in Eq.~\eqref{eq:generator} increments with size $L$ from $L(\mu{+}1)$ to $L(\nu{-}1)$. For example, a  $L{=}2$ pairwise Givens rotation would be
\be
\hat{U}^{(2)}_{\mu\nu}(\theta)=\exp \left[-i\theta\hat{X}_{2\mu-1}\hat{Y}_{2\mu}\hat{X}_{2\nu-1}\hat{X}_{2\nu}\hat{Z}_{2(\mu+1)}\hat{Z}_{2(\mu+2)}\ldots\hat{Z}_{2(\nu-1)}\right] \label{eq:apx_pair_givens}
\ee 
\noindent
where it can be easily decomposed as shown in Fig.~\ref{apx_fig1} using gate decomposition techniques from~\cite{whitfieldSimulationElectronicStructure2011}. Therefore, the Given rotation gate $\hat{U}^{(L)}_{\mu\nu}(\theta)$ used in this work is simply a Pauli-string rotation gate where its gate decomposition is a generalization of Fig.~\ref{apx_fig1} as shown in Fig. 3(b) of the main text.

% \newpage
\section{Derivation of the Two-Qubit Gate Depth Scaling of \texorpdfstring{$L$}{l}-wise Correlated Quantum Circuit Ansatz}
\label{apx:scaling}
To prepare the $L$-wise correlated ansatz state $\left|\Psi_{L}\right\rangle$ on a quantum computer under Jordan-Wigner mapping, we apply $\frac{d}{L}$ Clifford loaders $\hat{C}_L$ on a all-zero qubit state. Each Clifford loader $\hat{C}$ has two products of multi Givens rotation $D_L$ and $D^\dagger_L$. Each $D_L$ has $\left\lceil \log_2 \frac{N}{L} \right\rceil$ Givens rotation gate depth. Each Givens Rotation gate $U^{(L)}_{\mu\nu}$ contains two CNOT ladders $B^{(L)}_{\mu\nu}$ that acts on all $2L+\nu-\mu-1$ qubits in $\{L(\mu-1)+1,\ldots,L\mu\}$, $\{L(\mu+1),\ldots(\textrm{mod } L)\ldots,L(\nu-1)\}$ and $\{L(\nu-1)+1,\ldots,L\nu\}$. We employ the logarithmic-depth CNOT ladders as shown in Fig. 2 from the main text, thus each CNOT ladders $B^{(L)}_{\mu\nu}$ has a two-qubit gate depth of $\left\lceil\log_2 (2L+\nu-\mu-1)\right\rceil\approx \mathcal{O}(\log_2 N)$. Focusing on all Givens rotations $U^{(L)}_{1,2^s}$ that acts on the first qubit in every sublayer $s{\in}\{1,2,{\ldots},\left\lceil \log_2 \frac{N}{L} \right\rceil\}$ as shown in Fig.~3(a) from the main text, the overall two-qubit gate depth of the quantum circuit required to prepare $L$-wise correlated ansatz state $\left|\Psi_{L}\right\rangle$ is estimated to be,
\begin{align}
&\frac{2d}{L}\sum_{s=1}^{\left\lceil \log_2 \frac{N}{L} \right\rceil}2\left\lceil\log_2 (2L+2^s-1-1)\right\rceil\\
\le&\frac{4d}{L}\sum_{s=1}^{\left\lceil \log_2 \frac{N}{L} \right\rceil}\log_2 (2^{s+\log_2(L)})\\
=&\frac{4d}{L}\sum_{s=1}^{\left\lceil \log_2 \frac{N}{L} \right\rceil} \left(s+\log_2(L)\right)\\
=&\frac{2d}{L}\left(\left\lceil\log^2_2 \frac{N}{L}\right\rceil+ (1+2\log_2 L)\left\lceil\log_2 \frac{N}{L}\right\rceil \right).
\end{align}
% \newpage

\bibliography{main.bib}

\begin{thebibliography}{49}%
\makeatletter
\providecommand \@ifxundefined [1]{%
 \@ifx{#1\undefined}
}%
\providecommand \@ifnum [1]{%
 \ifnum #1\expandafter \@firstoftwo
 \else \expandafter \@secondoftwo
 \fi
}%
\providecommand \@ifx [1]{%
 \ifx #1\expandafter \@firstoftwo
 \else \expandafter \@secondoftwo
 \fi
}%
\providecommand \natexlab [1]{#1}%
\providecommand \enquote  [1]{``#1''}%
\providecommand \bibnamefont  [1]{#1}%
\providecommand \bibfnamefont [1]{#1}%
\providecommand \citenamefont [1]{#1}%
\providecommand \href@noop [0]{\@secondoftwo}%
\providecommand \href [0]{\begingroup \@sanitize@url \@href}%
\providecommand \@href[1]{\@@startlink{#1}\@@href}%
\providecommand \@@href[1]{\endgroup#1\@@endlink}%
\providecommand \@sanitize@url [0]{\catcode `\\12\catcode `\$12\catcode
  `\&12\catcode `\#12\catcode `\^12\catcode `\_12\catcode `\%12\relax}%
\providecommand \@@startlink[1]{}%
\providecommand \@@endlink[0]{}%
\providecommand \url  [0]{\begingroup\@sanitize@url \@url }%
\providecommand \@url [1]{\endgroup\@href {#1}{\urlprefix }}%
\providecommand \urlprefix  [0]{URL }%
\providecommand \Eprint [0]{\href }%
\providecommand \doibase [0]{https://doi.org/}%
\providecommand \selectlanguage [0]{\@gobble}%
\providecommand \bibinfo  [0]{\@secondoftwo}%
\providecommand \bibfield  [0]{\@secondoftwo}%
\providecommand \translation [1]{[#1]}%
\providecommand \BibitemOpen [0]{}%
\providecommand \bibitemStop [0]{}%
\providecommand \bibitemNoStop [0]{.\EOS\space}%
\providecommand \EOS [0]{\spacefactor3000\relax}%
\providecommand \BibitemShut  [1]{\csname bibitem#1\endcsname}%
\let\auto@bib@innerbib\@empty
%</preamble>
\bibitem [{\citenamefont {Bauer}\ \emph {et~al.}(2020)\citenamefont {Bauer},
  \citenamefont {Bravyi}, \citenamefont {Motta},\ and\ \citenamefont
  {Chan}}]{bauerQuantumAlgorithmsQuantum2020}%
  \BibitemOpen
  \bibfield  {author} {\bibinfo {author} {\bibfnamefont {B.}~\bibnamefont
  {Bauer}}, \bibinfo {author} {\bibfnamefont {S.}~\bibnamefont {Bravyi}},
  \bibinfo {author} {\bibfnamefont {M.}~\bibnamefont {Motta}},\ and\ \bibinfo
  {author} {\bibfnamefont {G.~K.-L.}\ \bibnamefont {Chan}},\ }\bibfield
  {title} {\bibinfo {title} {Quantum {{Algorithms}} for {{Quantum Chemistry}}
  and {{Quantum Materials Science}}},\ }\href
  {https://doi.org/10.1021/acs.chemrev.9b00829} {\bibfield  {journal} {\bibinfo
   {journal} {Chem. Rev.}\ }\textbf {\bibinfo {volume} {120}},\ \bibinfo
  {pages} {12685} (\bibinfo {year} {2020})}\BibitemShut {NoStop}%
\bibitem [{\citenamefont {Motta}\ and\ \citenamefont
  {Rice}(2022)}]{mottaEmergingQuantumComputing2022}%
  \BibitemOpen
  \bibfield  {author} {\bibinfo {author} {\bibfnamefont {M.}~\bibnamefont
  {Motta}}\ and\ \bibinfo {author} {\bibfnamefont {J.~E.}\ \bibnamefont
  {Rice}},\ }\bibfield  {title} {\bibinfo {title} {Emerging quantum computing
  algorithms for quantum chemistry},\ }\href
  {https://doi.org/10.1002/wcms.1580} {\bibfield  {journal} {\bibinfo
  {journal} {WIREs Computational Molecular Science}\ }\textbf {\bibinfo
  {volume} {12}},\ \bibinfo {pages} {e1580} (\bibinfo {year}
  {2022})}\BibitemShut {NoStop}%
\bibitem [{\citenamefont {Daley}\ \emph {et~al.}(2022)\citenamefont {Daley},
  \citenamefont {Bloch}, \citenamefont {Kokail}, \citenamefont {Flannigan},
  \citenamefont {Pearson}, \citenamefont {Troyer},\ and\ \citenamefont
  {Zoller}}]{daleyPracticalQuantumAdvantage2022}%
  \BibitemOpen
  \bibfield  {author} {\bibinfo {author} {\bibfnamefont {A.~J.}\ \bibnamefont
  {Daley}}, \bibinfo {author} {\bibfnamefont {I.}~\bibnamefont {Bloch}},
  \bibinfo {author} {\bibfnamefont {C.}~\bibnamefont {Kokail}}, \bibinfo
  {author} {\bibfnamefont {S.}~\bibnamefont {Flannigan}}, \bibinfo {author}
  {\bibfnamefont {N.}~\bibnamefont {Pearson}}, \bibinfo {author} {\bibfnamefont
  {M.}~\bibnamefont {Troyer}},\ and\ \bibinfo {author} {\bibfnamefont
  {P.}~\bibnamefont {Zoller}},\ }\bibfield  {title} {\bibinfo {title}
  {Practical quantum advantage in quantum simulation},\ }\href
  {https://doi.org/10.1038/s41586-022-04940-6} {\bibfield  {journal} {\bibinfo
  {journal} {Nature}\ }\textbf {\bibinfo {volume} {607}},\ \bibinfo {pages}
  {667} (\bibinfo {year} {2022})}\BibitemShut {NoStop}%
\bibitem [{\citenamefont {Abrams}\ and\ \citenamefont
  {Lloyd}(1997)}]{abramsSimulationManyBodyFermi1997}%
  \BibitemOpen
  \bibfield  {author} {\bibinfo {author} {\bibfnamefont {D.~S.}\ \bibnamefont
  {Abrams}}\ and\ \bibinfo {author} {\bibfnamefont {S.}~\bibnamefont {Lloyd}},\
  }\bibfield  {title} {\bibinfo {title} {Simulation of {{Many-Body Fermi
  Systems}} on a {{Universal Quantum Computer}}},\ }\href
  {https://doi.org/10.1103/PhysRevLett.79.2586} {\bibfield  {journal} {\bibinfo
   {journal} {Phys. Rev. Lett.}\ }\textbf {\bibinfo {volume} {79}},\ \bibinfo
  {pages} {2586} (\bibinfo {year} {1997})}\BibitemShut {NoStop}%
\bibitem [{\citenamefont {Abrams}\ and\ \citenamefont
  {Lloyd}(1999)}]{abramsQuantumAlgorithmProviding1999}%
  \BibitemOpen
  \bibfield  {author} {\bibinfo {author} {\bibfnamefont {D.~S.}\ \bibnamefont
  {Abrams}}\ and\ \bibinfo {author} {\bibfnamefont {S.}~\bibnamefont {Lloyd}},\
  }\bibfield  {title} {\bibinfo {title} {Quantum {{Algorithm Providing
  Exponential Speed Increase}} for {{Finding Eigenvalues}} and
  {{Eigenvectors}}},\ }\href {https://doi.org/10.1103/PhysRevLett.83.5162}
  {\bibfield  {journal} {\bibinfo  {journal} {Phys. Rev. Lett.}\ }\textbf
  {\bibinfo {volume} {83}},\ \bibinfo {pages} {5162} (\bibinfo {year}
  {1999})}\BibitemShut {NoStop}%
\bibitem [{\citenamefont {{Aspuru-Guzik}}\ \emph {et~al.}(2005)\citenamefont
  {{Aspuru-Guzik}}, \citenamefont {Dutoi}, \citenamefont {Love},\ and\
  \citenamefont {{Head-Gordon}}}]{aspuru-guzikSimulatedQuantumComputation2005}%
  \BibitemOpen
  \bibfield  {author} {\bibinfo {author} {\bibfnamefont {A.}~\bibnamefont
  {{Aspuru-Guzik}}}, \bibinfo {author} {\bibfnamefont {A.~D.}\ \bibnamefont
  {Dutoi}}, \bibinfo {author} {\bibfnamefont {P.~J.}\ \bibnamefont {Love}},\
  and\ \bibinfo {author} {\bibfnamefont {M.}~\bibnamefont {{Head-Gordon}}},\
  }\bibfield  {title} {\bibinfo {title} {Simulated {{Quantum Computation}} of
  {{Molecular Energies}}},\ }\href {https://doi.org/10.1126/science.1113479}
  {\bibfield  {journal} {\bibinfo  {journal} {Science}\ }\textbf {\bibinfo
  {volume} {309}},\ \bibinfo {pages} {1704} (\bibinfo {year}
  {2005})}\BibitemShut {NoStop}%
\bibitem [{\citenamefont {Lee}\ \emph {et~al.}(2022)\citenamefont {Lee},
  \citenamefont {Lee}, \citenamefont {Zhai}, \citenamefont {Tong},
  \citenamefont {Dalzell}, \citenamefont {Kumar}, \citenamefont {Helms},
  \citenamefont {Gray}, \citenamefont {Cui}, \citenamefont {Liu}, \citenamefont
  {Kastoryano}, \citenamefont {Babbush}, \citenamefont {Preskill},
  \citenamefont {Reichman}, \citenamefont {Campbell}, \citenamefont {Valeev},
  \citenamefont {Lin},\ and\ \citenamefont
  {Chan}}]{leeThereEvidenceExponential2022}%
  \BibitemOpen
  \bibfield  {author} {\bibinfo {author} {\bibfnamefont {S.}~\bibnamefont
  {Lee}}, \bibinfo {author} {\bibfnamefont {J.}~\bibnamefont {Lee}}, \bibinfo
  {author} {\bibfnamefont {H.}~\bibnamefont {Zhai}}, \bibinfo {author}
  {\bibfnamefont {Y.}~\bibnamefont {Tong}}, \bibinfo {author} {\bibfnamefont
  {A.~M.}\ \bibnamefont {Dalzell}}, \bibinfo {author} {\bibfnamefont
  {A.}~\bibnamefont {Kumar}}, \bibinfo {author} {\bibfnamefont
  {P.}~\bibnamefont {Helms}}, \bibinfo {author} {\bibfnamefont
  {J.}~\bibnamefont {Gray}}, \bibinfo {author} {\bibfnamefont {Z.-H.}\
  \bibnamefont {Cui}}, \bibinfo {author} {\bibfnamefont {W.}~\bibnamefont
  {Liu}}, \bibinfo {author} {\bibfnamefont {M.}~\bibnamefont {Kastoryano}},
  \bibinfo {author} {\bibfnamefont {R.}~\bibnamefont {Babbush}}, \bibinfo
  {author} {\bibfnamefont {J.}~\bibnamefont {Preskill}}, \bibinfo {author}
  {\bibfnamefont {D.~R.}\ \bibnamefont {Reichman}}, \bibinfo {author}
  {\bibfnamefont {E.~T.}\ \bibnamefont {Campbell}}, \bibinfo {author}
  {\bibfnamefont {E.~F.}\ \bibnamefont {Valeev}}, \bibinfo {author}
  {\bibfnamefont {L.}~\bibnamefont {Lin}},\ and\ \bibinfo {author}
  {\bibfnamefont {G.~K.-L.}\ \bibnamefont {Chan}},\ }\href
  {https://doi.org/10.48550/arXiv.2208.02199} {\bibinfo {title} {Is there
  evidence for exponential quantum advantage in quantum chemistry?}} (\bibinfo
  {year} {2022}),\ \Eprint {https://arxiv.org/abs/2208.02199}
  {arxiv:2208.02199} \BibitemShut {NoStop}%
\bibitem [{\citenamefont {Bharti}\ \emph {et~al.}(2022)\citenamefont {Bharti},
  \citenamefont {{Cervera-Lierta}}, \citenamefont {Kyaw}, \citenamefont {Haug},
  \citenamefont {{Alperin-Lea}}, \citenamefont {Anand}, \citenamefont
  {Degroote}, \citenamefont {Heimonen}, \citenamefont {Kottmann}, \citenamefont
  {Menke}, \citenamefont {Mok}, \citenamefont {Sim}, \citenamefont {Kwek},\
  and\ \citenamefont
  {{Aspuru-Guzik}}}]{bhartiNoisyIntermediatescaleQuantum2022}%
  \BibitemOpen
  \bibfield  {author} {\bibinfo {author} {\bibfnamefont {K.}~\bibnamefont
  {Bharti}}, \bibinfo {author} {\bibfnamefont {A.}~\bibnamefont
  {{Cervera-Lierta}}}, \bibinfo {author} {\bibfnamefont {T.~H.}\ \bibnamefont
  {Kyaw}}, \bibinfo {author} {\bibfnamefont {T.}~\bibnamefont {Haug}}, \bibinfo
  {author} {\bibfnamefont {S.}~\bibnamefont {{Alperin-Lea}}}, \bibinfo {author}
  {\bibfnamefont {A.}~\bibnamefont {Anand}}, \bibinfo {author} {\bibfnamefont
  {M.}~\bibnamefont {Degroote}}, \bibinfo {author} {\bibfnamefont
  {H.}~\bibnamefont {Heimonen}}, \bibinfo {author} {\bibfnamefont {J.~S.}\
  \bibnamefont {Kottmann}}, \bibinfo {author} {\bibfnamefont {T.}~\bibnamefont
  {Menke}}, \bibinfo {author} {\bibfnamefont {W.-K.}\ \bibnamefont {Mok}},
  \bibinfo {author} {\bibfnamefont {S.}~\bibnamefont {Sim}}, \bibinfo {author}
  {\bibfnamefont {L.-C.}\ \bibnamefont {Kwek}},\ and\ \bibinfo {author}
  {\bibfnamefont {A.}~\bibnamefont {{Aspuru-Guzik}}},\ }\bibfield  {title}
  {\bibinfo {title} {Noisy intermediate-scale quantum algorithms},\ }\href
  {https://doi.org/10.1103/RevModPhys.94.015004} {\bibfield  {journal}
  {\bibinfo  {journal} {Rev. Mod. Phys.}\ }\textbf {\bibinfo {volume} {94}},\
  \bibinfo {pages} {015004} (\bibinfo {year} {2022})}\BibitemShut {NoStop}%
\bibitem [{\citenamefont {Jordan}\ and\ \citenamefont
  {Wigner}(1928)}]{jordanUeberPaulischeAequivalenzverbot1928}%
  \BibitemOpen
  \bibfield  {author} {\bibinfo {author} {\bibfnamefont {P.}~\bibnamefont
  {Jordan}}\ and\ \bibinfo {author} {\bibfnamefont {E.}~\bibnamefont
  {Wigner}},\ }\bibfield  {title} {\bibinfo {title} {{\"Uber das Paulische
  \"Aquivalenzverbot}},\ }\href {https://doi.org/10.1007/BF01331938} {\bibfield
   {journal} {\bibinfo  {journal} {Z. Physik}\ }\textbf {\bibinfo {volume}
  {47}},\ \bibinfo {pages} {631} (\bibinfo {year} {1928})}\BibitemShut
  {NoStop}%
\bibitem [{\citenamefont {Kandala}\ \emph {et~al.}(2017)\citenamefont
  {Kandala}, \citenamefont {Mezzacapo}, \citenamefont {Temme}, \citenamefont
  {Takita}, \citenamefont {Brink}, \citenamefont {Chow},\ and\ \citenamefont
  {Gambetta}}]{kandalaHardwareefficientVariationalQuantum2017}%
  \BibitemOpen
  \bibfield  {author} {\bibinfo {author} {\bibfnamefont {A.}~\bibnamefont
  {Kandala}}, \bibinfo {author} {\bibfnamefont {A.}~\bibnamefont {Mezzacapo}},
  \bibinfo {author} {\bibfnamefont {K.}~\bibnamefont {Temme}}, \bibinfo
  {author} {\bibfnamefont {M.}~\bibnamefont {Takita}}, \bibinfo {author}
  {\bibfnamefont {M.}~\bibnamefont {Brink}}, \bibinfo {author} {\bibfnamefont
  {J.~M.}\ \bibnamefont {Chow}},\ and\ \bibinfo {author} {\bibfnamefont
  {J.~M.}\ \bibnamefont {Gambetta}},\ }\bibfield  {title} {\bibinfo {title}
  {Hardware-efficient variational quantum eigensolver for small molecules and
  quantum magnets},\ }\href {https://doi.org/10.1038/nature23879} {\bibfield
  {journal} {\bibinfo  {journal} {Nature}\ }\textbf {\bibinfo {volume} {549}},\
  \bibinfo {pages} {242} (\bibinfo {year} {2017})}\BibitemShut {NoStop}%
\bibitem [{\citenamefont {Bittel}\ and\ \citenamefont
  {Kliesch}(2021)}]{bittelTrainingVariationalQuantum2021}%
  \BibitemOpen
  \bibfield  {author} {\bibinfo {author} {\bibfnamefont {L.}~\bibnamefont
  {Bittel}}\ and\ \bibinfo {author} {\bibfnamefont {M.}~\bibnamefont
  {Kliesch}},\ }\bibfield  {title} {\bibinfo {title} {Training {{Variational
  Quantum Algorithms Is NP-Hard}}},\ }\href
  {https://doi.org/10.1103/PhysRevLett.127.120502} {\bibfield  {journal}
  {\bibinfo  {journal} {Phys. Rev. Lett.}\ }\textbf {\bibinfo {volume} {127}},\
  \bibinfo {pages} {120502} (\bibinfo {year} {2021})}\BibitemShut {NoStop}%
\bibitem [{\citenamefont {Tilly}\ \emph {et~al.}(2022)\citenamefont {Tilly},
  \citenamefont {Chen}, \citenamefont {Cao}, \citenamefont {Picozzi},
  \citenamefont {Setia}, \citenamefont {Li}, \citenamefont {Grant},
  \citenamefont {Wossnig}, \citenamefont {Rungger}, \citenamefont {Booth},\
  and\ \citenamefont {Tennyson}}]{tillyVariationalQuantumEigensolver2022}%
  \BibitemOpen
  \bibfield  {author} {\bibinfo {author} {\bibfnamefont {J.}~\bibnamefont
  {Tilly}}, \bibinfo {author} {\bibfnamefont {H.}~\bibnamefont {Chen}},
  \bibinfo {author} {\bibfnamefont {S.}~\bibnamefont {Cao}}, \bibinfo {author}
  {\bibfnamefont {D.}~\bibnamefont {Picozzi}}, \bibinfo {author} {\bibfnamefont
  {K.}~\bibnamefont {Setia}}, \bibinfo {author} {\bibfnamefont
  {Y.}~\bibnamefont {Li}}, \bibinfo {author} {\bibfnamefont {E.}~\bibnamefont
  {Grant}}, \bibinfo {author} {\bibfnamefont {L.}~\bibnamefont {Wossnig}},
  \bibinfo {author} {\bibfnamefont {I.}~\bibnamefont {Rungger}}, \bibinfo
  {author} {\bibfnamefont {G.~H.}\ \bibnamefont {Booth}},\ and\ \bibinfo
  {author} {\bibfnamefont {J.}~\bibnamefont {Tennyson}},\ }\bibfield  {title}
  {\bibinfo {title} {The {{Variational Quantum Eigensolver}}: {{A}} review of
  methods and best practices},\ }\href
  {https://doi.org/10.1016/j.physrep.2022.08.003} {\bibfield  {journal}
  {\bibinfo  {journal} {Physics Reports}\ }\bibinfo {series} {The {{Variational
  Quantum Eigensolver}}: A Review of Methods and Best Practices},\ \textbf
  {\bibinfo {volume} {986}},\ \bibinfo {pages} {1} (\bibinfo {year}
  {2022})}\BibitemShut {NoStop}%
\bibitem [{\citenamefont {Wecker}\ \emph {et~al.}(2015)\citenamefont {Wecker},
  \citenamefont {Hastings}, \citenamefont {Wiebe}, \citenamefont {Clark},
  \citenamefont {Nayak},\ and\ \citenamefont
  {Troyer}}]{weckerSolvingStronglyCorrelated2015}%
  \BibitemOpen
  \bibfield  {author} {\bibinfo {author} {\bibfnamefont {D.}~\bibnamefont
  {Wecker}}, \bibinfo {author} {\bibfnamefont {M.~B.}\ \bibnamefont
  {Hastings}}, \bibinfo {author} {\bibfnamefont {N.}~\bibnamefont {Wiebe}},
  \bibinfo {author} {\bibfnamefont {B.~K.}\ \bibnamefont {Clark}}, \bibinfo
  {author} {\bibfnamefont {C.}~\bibnamefont {Nayak}},\ and\ \bibinfo {author}
  {\bibfnamefont {M.}~\bibnamefont {Troyer}},\ }\bibfield  {title} {\bibinfo
  {title} {Solving strongly correlated electron models on a quantum computer},\
  }\href {https://doi.org/10.1103/PhysRevA.92.062318} {\bibfield  {journal}
  {\bibinfo  {journal} {Phys. Rev. A}\ }\textbf {\bibinfo {volume} {92}},\
  \bibinfo {pages} {062318} (\bibinfo {year} {2015})}\BibitemShut {NoStop}%
\bibitem [{\citenamefont {Kivlichan}\ \emph {et~al.}(2018)\citenamefont
  {Kivlichan}, \citenamefont {McClean}, \citenamefont {Wiebe}, \citenamefont
  {Gidney}, \citenamefont {{Aspuru-Guzik}}, \citenamefont {Chan},\ and\
  \citenamefont {Babbush}}]{kivlichanQuantumSimulationElectronic2018}%
  \BibitemOpen
  \bibfield  {author} {\bibinfo {author} {\bibfnamefont {I.~D.}\ \bibnamefont
  {Kivlichan}}, \bibinfo {author} {\bibfnamefont {J.}~\bibnamefont {McClean}},
  \bibinfo {author} {\bibfnamefont {N.}~\bibnamefont {Wiebe}}, \bibinfo
  {author} {\bibfnamefont {C.}~\bibnamefont {Gidney}}, \bibinfo {author}
  {\bibfnamefont {A.}~\bibnamefont {{Aspuru-Guzik}}}, \bibinfo {author}
  {\bibfnamefont {G.~K.-L.}\ \bibnamefont {Chan}},\ and\ \bibinfo {author}
  {\bibfnamefont {R.}~\bibnamefont {Babbush}},\ }\bibfield  {title} {\bibinfo
  {title} {Quantum {{Simulation}} of {{Electronic Structure}} with {{Linear
  Depth}} and {{Connectivity}}},\ }\href
  {https://doi.org/10.1103/PhysRevLett.120.110501} {\bibfield  {journal}
  {\bibinfo  {journal} {Phys. Rev. Lett.}\ }\textbf {\bibinfo {volume} {120}},\
  \bibinfo {pages} {110501} (\bibinfo {year} {2018})}\BibitemShut {NoStop}%
\bibitem [{\citenamefont {Arute}\ \emph {et~al.}(2020)\citenamefont {Arute},
  \citenamefont {Arya}, \citenamefont {Babbush}, \citenamefont {Bacon},
  \citenamefont {Bardin}, \citenamefont {Barends}, \citenamefont {Boixo},
  \citenamefont {Broughton}, \citenamefont {Buckley}, \citenamefont {Buell},
  \citenamefont {Burkett}, \citenamefont {Bushnell}, \citenamefont {Chen},
  \citenamefont {Chen}, \citenamefont {Chiaro}, \citenamefont {Collins},
  \citenamefont {Courtney}, \citenamefont {Demura}, \citenamefont {Dunsworth},
  \citenamefont {Farhi}, \citenamefont {Fowler}, \citenamefont {Foxen},
  \citenamefont {Gidney}, \citenamefont {Giustina}, \citenamefont {Graff},
  \citenamefont {Habegger}, \citenamefont {Harrigan}, \citenamefont {Ho},
  \citenamefont {Hong}, \citenamefont {Huang}, \citenamefont {Huggins},
  \citenamefont {Ioffe}, \citenamefont {Isakov}, \citenamefont {Jeffrey},
  \citenamefont {Jiang}, \citenamefont {Jones}, \citenamefont {Kafri},
  \citenamefont {Kechedzhi}, \citenamefont {Kelly}, \citenamefont {Kim},
  \citenamefont {Klimov}, \citenamefont {Korotkov}, \citenamefont {Kostritsa},
  \citenamefont {Landhuis}, \citenamefont {Laptev}, \citenamefont {Lindmark},
  \citenamefont {Lucero}, \citenamefont {Martin}, \citenamefont {Martinis},
  \citenamefont {McClean}, \citenamefont {McEwen}, \citenamefont {Megrant},
  \citenamefont {Mi}, \citenamefont {Mohseni}, \citenamefont {Mruczkiewicz},
  \citenamefont {Mutus}, \citenamefont {Naaman}, \citenamefont {Neeley},
  \citenamefont {Neill}, \citenamefont {Neven}, \citenamefont {Niu},
  \citenamefont {O'Brien}, \citenamefont {Ostby}, \citenamefont {Petukhov},
  \citenamefont {Putterman}, \citenamefont {Quintana}, \citenamefont {Roushan},
  \citenamefont {Rubin}, \citenamefont {Sank}, \citenamefont {Satzinger},
  \citenamefont {Smelyanskiy}, \citenamefont {Strain}, \citenamefont {Sung},
  \citenamefont {Szalay}, \citenamefont {Takeshita}, \citenamefont
  {Vainsencher}, \citenamefont {White}, \citenamefont {Wiebe}, \citenamefont
  {Yao}, \citenamefont {Yeh},\ and\ \citenamefont
  {Zalcman}}]{aruteHartreeFockSuperconductingQubit2020}%
  \BibitemOpen
  \bibfield  {author} {\bibinfo {author} {\bibfnamefont {F.}~\bibnamefont
  {Arute}}, \bibinfo {author} {\bibfnamefont {K.}~\bibnamefont {Arya}},
  \bibinfo {author} {\bibfnamefont {R.}~\bibnamefont {Babbush}}, \bibinfo
  {author} {\bibfnamefont {D.}~\bibnamefont {Bacon}}, \bibinfo {author}
  {\bibfnamefont {J.~C.}\ \bibnamefont {Bardin}}, \bibinfo {author}
  {\bibfnamefont {R.}~\bibnamefont {Barends}}, \bibinfo {author} {\bibfnamefont
  {S.}~\bibnamefont {Boixo}}, \bibinfo {author} {\bibfnamefont
  {M.}~\bibnamefont {Broughton}}, \bibinfo {author} {\bibfnamefont {B.~B.}\
  \bibnamefont {Buckley}}, \bibinfo {author} {\bibfnamefont {D.~A.}\
  \bibnamefont {Buell}}, \bibinfo {author} {\bibfnamefont {B.}~\bibnamefont
  {Burkett}}, \bibinfo {author} {\bibfnamefont {N.}~\bibnamefont {Bushnell}},
  \bibinfo {author} {\bibfnamefont {Y.}~\bibnamefont {Chen}}, \bibinfo {author}
  {\bibfnamefont {Z.}~\bibnamefont {Chen}}, \bibinfo {author} {\bibfnamefont
  {B.}~\bibnamefont {Chiaro}}, \bibinfo {author} {\bibfnamefont
  {R.}~\bibnamefont {Collins}}, \bibinfo {author} {\bibfnamefont
  {W.}~\bibnamefont {Courtney}}, \bibinfo {author} {\bibfnamefont
  {S.}~\bibnamefont {Demura}}, \bibinfo {author} {\bibfnamefont
  {A.}~\bibnamefont {Dunsworth}}, \bibinfo {author} {\bibfnamefont
  {E.}~\bibnamefont {Farhi}}, \bibinfo {author} {\bibfnamefont
  {A.}~\bibnamefont {Fowler}}, \bibinfo {author} {\bibfnamefont
  {B.}~\bibnamefont {Foxen}}, \bibinfo {author} {\bibfnamefont
  {C.}~\bibnamefont {Gidney}}, \bibinfo {author} {\bibfnamefont
  {M.}~\bibnamefont {Giustina}}, \bibinfo {author} {\bibfnamefont
  {R.}~\bibnamefont {Graff}}, \bibinfo {author} {\bibfnamefont
  {S.}~\bibnamefont {Habegger}}, \bibinfo {author} {\bibfnamefont {M.~P.}\
  \bibnamefont {Harrigan}}, \bibinfo {author} {\bibfnamefont {A.}~\bibnamefont
  {Ho}}, \bibinfo {author} {\bibfnamefont {S.}~\bibnamefont {Hong}}, \bibinfo
  {author} {\bibfnamefont {T.}~\bibnamefont {Huang}}, \bibinfo {author}
  {\bibfnamefont {W.~J.}\ \bibnamefont {Huggins}}, \bibinfo {author}
  {\bibfnamefont {L.}~\bibnamefont {Ioffe}}, \bibinfo {author} {\bibfnamefont
  {S.~V.}\ \bibnamefont {Isakov}}, \bibinfo {author} {\bibfnamefont
  {E.}~\bibnamefont {Jeffrey}}, \bibinfo {author} {\bibfnamefont
  {Z.}~\bibnamefont {Jiang}}, \bibinfo {author} {\bibfnamefont
  {C.}~\bibnamefont {Jones}}, \bibinfo {author} {\bibfnamefont
  {D.}~\bibnamefont {Kafri}}, \bibinfo {author} {\bibfnamefont
  {K.}~\bibnamefont {Kechedzhi}}, \bibinfo {author} {\bibfnamefont
  {J.}~\bibnamefont {Kelly}}, \bibinfo {author} {\bibfnamefont
  {S.}~\bibnamefont {Kim}}, \bibinfo {author} {\bibfnamefont {P.~V.}\
  \bibnamefont {Klimov}}, \bibinfo {author} {\bibfnamefont {A.}~\bibnamefont
  {Korotkov}}, \bibinfo {author} {\bibfnamefont {F.}~\bibnamefont {Kostritsa}},
  \bibinfo {author} {\bibfnamefont {D.}~\bibnamefont {Landhuis}}, \bibinfo
  {author} {\bibfnamefont {P.}~\bibnamefont {Laptev}}, \bibinfo {author}
  {\bibfnamefont {M.}~\bibnamefont {Lindmark}}, \bibinfo {author}
  {\bibfnamefont {E.}~\bibnamefont {Lucero}}, \bibinfo {author} {\bibfnamefont
  {O.}~\bibnamefont {Martin}}, \bibinfo {author} {\bibfnamefont {J.~M.}\
  \bibnamefont {Martinis}}, \bibinfo {author} {\bibfnamefont {J.~R.}\
  \bibnamefont {McClean}}, \bibinfo {author} {\bibfnamefont {M.}~\bibnamefont
  {McEwen}}, \bibinfo {author} {\bibfnamefont {A.}~\bibnamefont {Megrant}},
  \bibinfo {author} {\bibfnamefont {X.}~\bibnamefont {Mi}}, \bibinfo {author}
  {\bibfnamefont {M.}~\bibnamefont {Mohseni}}, \bibinfo {author} {\bibfnamefont
  {W.}~\bibnamefont {Mruczkiewicz}}, \bibinfo {author} {\bibfnamefont
  {J.}~\bibnamefont {Mutus}}, \bibinfo {author} {\bibfnamefont
  {O.}~\bibnamefont {Naaman}}, \bibinfo {author} {\bibfnamefont
  {M.}~\bibnamefont {Neeley}}, \bibinfo {author} {\bibfnamefont
  {C.}~\bibnamefont {Neill}}, \bibinfo {author} {\bibfnamefont
  {H.}~\bibnamefont {Neven}}, \bibinfo {author} {\bibfnamefont {M.~Y.}\
  \bibnamefont {Niu}}, \bibinfo {author} {\bibfnamefont {T.~E.}\ \bibnamefont
  {O'Brien}}, \bibinfo {author} {\bibfnamefont {E.}~\bibnamefont {Ostby}},
  \bibinfo {author} {\bibfnamefont {A.}~\bibnamefont {Petukhov}}, \bibinfo
  {author} {\bibfnamefont {H.}~\bibnamefont {Putterman}}, \bibinfo {author}
  {\bibfnamefont {C.}~\bibnamefont {Quintana}}, \bibinfo {author}
  {\bibfnamefont {P.}~\bibnamefont {Roushan}}, \bibinfo {author} {\bibfnamefont
  {N.~C.}\ \bibnamefont {Rubin}}, \bibinfo {author} {\bibfnamefont
  {D.}~\bibnamefont {Sank}}, \bibinfo {author} {\bibfnamefont {K.~J.}\
  \bibnamefont {Satzinger}}, \bibinfo {author} {\bibfnamefont {V.}~\bibnamefont
  {Smelyanskiy}}, \bibinfo {author} {\bibfnamefont {D.}~\bibnamefont {Strain}},
  \bibinfo {author} {\bibfnamefont {K.~J.}\ \bibnamefont {Sung}}, \bibinfo
  {author} {\bibfnamefont {M.}~\bibnamefont {Szalay}}, \bibinfo {author}
  {\bibfnamefont {T.~Y.}\ \bibnamefont {Takeshita}}, \bibinfo {author}
  {\bibfnamefont {A.}~\bibnamefont {Vainsencher}}, \bibinfo {author}
  {\bibfnamefont {T.}~\bibnamefont {White}}, \bibinfo {author} {\bibfnamefont
  {N.}~\bibnamefont {Wiebe}}, \bibinfo {author} {\bibfnamefont {Z.~J.}\
  \bibnamefont {Yao}}, \bibinfo {author} {\bibfnamefont {P.}~\bibnamefont
  {Yeh}},\ and\ \bibinfo {author} {\bibfnamefont {A.}~\bibnamefont {Zalcman}},\
  }\bibfield  {title} {\bibinfo {title} {Hartree-{{Fock}} on a superconducting
  qubit quantum computer},\ }\href {https://doi.org/10.1126/science.abb9811}
  {\bibfield  {journal} {\bibinfo  {journal} {Science}\ }\textbf {\bibinfo
  {volume} {369}},\ \bibinfo {pages} {1084} (\bibinfo {year}
  {2020})}\BibitemShut {NoStop}%
\bibitem [{\citenamefont {Anand}\ \emph {et~al.}(2022)\citenamefont {Anand},
  \citenamefont {Schleich}, \citenamefont {{Alperin-Lea}}, \citenamefont
  {Jensen}, \citenamefont {Sim}, \citenamefont {{D{\'i}az-Tinoco}},
  \citenamefont {Kottmann}, \citenamefont {Degroote}, \citenamefont
  {Izmaylov},\ and\ \citenamefont
  {{Aspuru-Guzik}}}]{anandQuantumComputingView2022}%
  \BibitemOpen
  \bibfield  {author} {\bibinfo {author} {\bibfnamefont {A.}~\bibnamefont
  {Anand}}, \bibinfo {author} {\bibfnamefont {P.}~\bibnamefont {Schleich}},
  \bibinfo {author} {\bibfnamefont {S.}~\bibnamefont {{Alperin-Lea}}}, \bibinfo
  {author} {\bibfnamefont {P.~W.~K.}\ \bibnamefont {Jensen}}, \bibinfo {author}
  {\bibfnamefont {S.}~\bibnamefont {Sim}}, \bibinfo {author} {\bibfnamefont
  {M.}~\bibnamefont {{D{\'i}az-Tinoco}}}, \bibinfo {author} {\bibfnamefont
  {J.~S.}\ \bibnamefont {Kottmann}}, \bibinfo {author} {\bibfnamefont
  {M.}~\bibnamefont {Degroote}}, \bibinfo {author} {\bibfnamefont {A.~F.}\
  \bibnamefont {Izmaylov}},\ and\ \bibinfo {author} {\bibfnamefont
  {A.}~\bibnamefont {{Aspuru-Guzik}}},\ }\bibfield  {title} {\bibinfo {title}
  {A quantum computing view on unitary coupled cluster theory},\ }\href
  {https://doi.org/10.1039/D1CS00932J} {\bibfield  {journal} {\bibinfo
  {journal} {Chem. Soc. Rev.}\ }\textbf {\bibinfo {volume} {51}},\ \bibinfo
  {pages} {1659} (\bibinfo {year} {2022})}\BibitemShut {NoStop}%
\bibitem [{\citenamefont {Evangelista}\ \emph {et~al.}(2019)\citenamefont
  {Evangelista}, \citenamefont {Chan},\ and\ \citenamefont
  {Scuseria}}]{evangelistaExactParameterizationFermionic2019}%
  \BibitemOpen
  \bibfield  {author} {\bibinfo {author} {\bibfnamefont {F.~A.}\ \bibnamefont
  {Evangelista}}, \bibinfo {author} {\bibfnamefont {G.~K.-L.}\ \bibnamefont
  {Chan}},\ and\ \bibinfo {author} {\bibfnamefont {G.~E.}\ \bibnamefont
  {Scuseria}},\ }\bibfield  {title} {\bibinfo {title} {Exact parameterization
  of fermionic wave functions via unitary coupled cluster theory},\ }\href
  {https://doi.org/10.1063/1.5133059} {\bibfield  {journal} {\bibinfo
  {journal} {J. Chem. Phys.}\ }\textbf {\bibinfo {volume} {151}},\ \bibinfo
  {pages} {244112} (\bibinfo {year} {2019})}\BibitemShut {NoStop}%
\bibitem [{\citenamefont {Wang}\ \emph {et~al.}(2021)\citenamefont {Wang},
  \citenamefont {Li}, \citenamefont {Monroe},\ and\ \citenamefont
  {Nam}}]{wangResourceOptimizedFermionicLocalHamiltonian2021}%
  \BibitemOpen
  \bibfield  {author} {\bibinfo {author} {\bibfnamefont {Q.}~\bibnamefont
  {Wang}}, \bibinfo {author} {\bibfnamefont {M.}~\bibnamefont {Li}}, \bibinfo
  {author} {\bibfnamefont {C.}~\bibnamefont {Monroe}},\ and\ \bibinfo {author}
  {\bibfnamefont {Y.}~\bibnamefont {Nam}},\ }\bibfield  {title} {\bibinfo
  {title} {Resource-{{Optimized Fermionic Local-Hamiltonian Simulation}} on a
  {{Quantum Computer}} for {{Quantum Chemistry}}},\ }\href
  {https://doi.org/10.22331/q-2021-07-26-509} {\bibfield  {journal} {\bibinfo
  {journal} {Quantum}\ }\textbf {\bibinfo {volume} {5}},\ \bibinfo {pages}
  {509} (\bibinfo {year} {2021})}\BibitemShut {NoStop}%
\bibitem [{\citenamefont {Kottmann}\ and\ \citenamefont
  {{Aspuru-Guzik}}(2022)}]{kottmannOptimizedLowdepthQuantum2022}%
  \BibitemOpen
  \bibfield  {author} {\bibinfo {author} {\bibfnamefont {J.~S.}\ \bibnamefont
  {Kottmann}}\ and\ \bibinfo {author} {\bibfnamefont {A.}~\bibnamefont
  {{Aspuru-Guzik}}},\ }\bibfield  {title} {\bibinfo {title} {Optimized
  low-depth quantum circuits for molecular electronic structure using a
  separable-pair approximation},\ }\href
  {https://doi.org/10.1103/PhysRevA.105.032449} {\bibfield  {journal} {\bibinfo
   {journal} {Phys. Rev. A}\ }\textbf {\bibinfo {volume} {105}},\ \bibinfo
  {pages} {032449} (\bibinfo {year} {2022})}\BibitemShut {NoStop}%
\bibitem [{\citenamefont {Tang}\ \emph {et~al.}(2021)\citenamefont {Tang},
  \citenamefont {Shkolnikov}, \citenamefont {Barron}, \citenamefont {Grimsley},
  \citenamefont {Mayhall}, \citenamefont {Barnes},\ and\ \citenamefont
  {Economou}}]{tangQubitADAPTVQEAdaptiveAlgorithm2021}%
  \BibitemOpen
  \bibfield  {author} {\bibinfo {author} {\bibfnamefont {H.~L.}\ \bibnamefont
  {Tang}}, \bibinfo {author} {\bibfnamefont {V.~O.}\ \bibnamefont
  {Shkolnikov}}, \bibinfo {author} {\bibfnamefont {G.~S.}\ \bibnamefont
  {Barron}}, \bibinfo {author} {\bibfnamefont {H.~R.}\ \bibnamefont
  {Grimsley}}, \bibinfo {author} {\bibfnamefont {N.~J.}\ \bibnamefont
  {Mayhall}}, \bibinfo {author} {\bibfnamefont {E.}~\bibnamefont {Barnes}},\
  and\ \bibinfo {author} {\bibfnamefont {S.~E.}\ \bibnamefont {Economou}},\
  }\bibfield  {title} {\bibinfo {title} {Qubit-{{ADAPT-VQE}}: {{An Adaptive
  Algorithm}} for {{Constructing Hardware-Efficient Ans}}\textbackslash
  "\{a\}tze on a {{Quantum Processor}}},\ }\href
  {https://doi.org/10.1103/PRXQuantum.2.020310} {\bibfield  {journal} {\bibinfo
   {journal} {PRX Quantum}\ }\textbf {\bibinfo {volume} {2}},\ \bibinfo {pages}
  {020310} (\bibinfo {year} {2021})}\BibitemShut {NoStop}%
\bibitem [{\citenamefont {O'Brien}\ \emph {et~al.}(2022)\citenamefont
  {O'Brien}, \citenamefont {Anselmetti}, \citenamefont {Gkritsis},
  \citenamefont {Elfving}, \citenamefont {Polla}, \citenamefont {Huggins},
  \citenamefont {Oumarou}, \citenamefont {Kechedzhi}, \citenamefont {Abanin},
  \citenamefont {Acharya}, \citenamefont {Aleiner}, \citenamefont {Allen},
  \citenamefont {Andersen}, \citenamefont {Anderson}, \citenamefont {Ansmann},
  \citenamefont {Arute}, \citenamefont {Arya}, \citenamefont {Asfaw},
  \citenamefont {Atalaya}, \citenamefont {Bacon}, \citenamefont {Bardin},
  \citenamefont {Bengtsson}, \citenamefont {Boixo}, \citenamefont {Bortoli},
  \citenamefont {Bourassa}, \citenamefont {Bovaird}, \citenamefont {Brill},
  \citenamefont {Broughton}, \citenamefont {Buckley}, \citenamefont {Buell},
  \citenamefont {Burger}, \citenamefont {Burkett}, \citenamefont {Bushnell},
  \citenamefont {Campero}, \citenamefont {Chen}, \citenamefont {Chen},
  \citenamefont {Chiaro}, \citenamefont {Chik}, \citenamefont {Cogan},
  \citenamefont {Collins}, \citenamefont {Conner}, \citenamefont {Courtney},
  \citenamefont {Crook}, \citenamefont {Curtin}, \citenamefont {Debroy},
  \citenamefont {Demura}, \citenamefont {Drozdov}, \citenamefont {Dunsworth},
  \citenamefont {Erickson}, \citenamefont {Faoro}, \citenamefont {Farhi},
  \citenamefont {Fatemi}, \citenamefont {Ferreira}, \citenamefont {Burgos},
  \citenamefont {Forati}, \citenamefont {Fowler}, \citenamefont {Foxen},
  \citenamefont {Giang}, \citenamefont {Gidney}, \citenamefont {Gilboa},
  \citenamefont {Giustina}, \citenamefont {Gosula}, \citenamefont {Dau},
  \citenamefont {Gross}, \citenamefont {Habegger}, \citenamefont {Hamilton},
  \citenamefont {Hansen}, \citenamefont {Harrigan}, \citenamefont {Harrington},
  \citenamefont {Heu}, \citenamefont {Hilton}, \citenamefont {Hoffmann},
  \citenamefont {Hong}, \citenamefont {Huang}, \citenamefont {Huff},
  \citenamefont {Ioffe}, \citenamefont {Isakov}, \citenamefont {Iveland},
  \citenamefont {Jeffrey}, \citenamefont {Jiang}, \citenamefont {Jones},
  \citenamefont {Juhas}, \citenamefont {Kafri}, \citenamefont {Kelly},
  \citenamefont {Khattar}, \citenamefont {Khezri}, \citenamefont
  {Kieferov{\'a}}, \citenamefont {Kim}, \citenamefont {Klimov}, \citenamefont
  {Klots}, \citenamefont {Kothari}, \citenamefont {Korotkov}, \citenamefont
  {Kostritsa}, \citenamefont {Kreikebaum}, \citenamefont {Landhuis},
  \citenamefont {Laptev}, \citenamefont {Lau}, \citenamefont {Laws},
  \citenamefont {Lee}, \citenamefont {Lee}, \citenamefont {Lester},
  \citenamefont {Lill}, \citenamefont {Liu}, \citenamefont {Livingston},
  \citenamefont {Locharla}, \citenamefont {Lucero}, \citenamefont {Malone},
  \citenamefont {Mandra}, \citenamefont {Martin}, \citenamefont {Martin},
  \citenamefont {McClean}, \citenamefont {McCourt}, \citenamefont {McEwen},
  \citenamefont {Megrant}, \citenamefont {Mi}, \citenamefont {Mieszala},
  \citenamefont {Miao}, \citenamefont {Mohseni}, \citenamefont {Montazeri},
  \citenamefont {Morvan}, \citenamefont {Movassagh}, \citenamefont
  {Mruczkiewicz}, \citenamefont {Naaman}, \citenamefont {Neeley}, \citenamefont
  {Neill}, \citenamefont {Nersisyan}, \citenamefont {Neven}, \citenamefont
  {Newman}, \citenamefont {Ng}, \citenamefont {Nguyen}, \citenamefont {Nguyen},
  \citenamefont {Niu}, \citenamefont {Omonije}, \citenamefont {Opremcak},
  \citenamefont {Petukhov}, \citenamefont {Potter}, \citenamefont {Pryadko},
  \citenamefont {Quintana}, \citenamefont {Rocque}, \citenamefont {Roushan},
  \citenamefont {Saei}, \citenamefont {Sank}, \citenamefont {Sankaragomathi},
  \citenamefont {Satzinger}, \citenamefont {Schurkus}, \citenamefont
  {Schuster}, \citenamefont {Shearn}, \citenamefont {Shorter}, \citenamefont
  {Shutty}, \citenamefont {Shvarts}, \citenamefont {Skruzny}, \citenamefont
  {Smelyanskiy}, \citenamefont {Smith}, \citenamefont {Somma}, \citenamefont
  {Sterling}, \citenamefont {Strain}, \citenamefont {Szalay}, \citenamefont
  {Thor}, \citenamefont {Torres}, \citenamefont {Vidal}, \citenamefont
  {Villalonga}, \citenamefont {Heidweiller}, \citenamefont {White},
  \citenamefont {Woo}, \citenamefont {Xing}, \citenamefont {Yao}, \citenamefont
  {Yeh}, \citenamefont {Yoo}, \citenamefont {Young}, \citenamefont {Zalcman},
  \citenamefont {Zhang}, \citenamefont {Zhu}, \citenamefont {Zobrist},
  \citenamefont {Gogolin}, \citenamefont {Babbush},\ and\ \citenamefont
  {Rubin}}]{obrienPurificationbasedQuantumError2022}%
  \BibitemOpen
  \bibfield  {author} {\bibinfo {author} {\bibfnamefont {T.~E.}\ \bibnamefont
  {O'Brien}}, \bibinfo {author} {\bibfnamefont {G.}~\bibnamefont {Anselmetti}},
  \bibinfo {author} {\bibfnamefont {F.}~\bibnamefont {Gkritsis}}, \bibinfo
  {author} {\bibfnamefont {V.~E.}\ \bibnamefont {Elfving}}, \bibinfo {author}
  {\bibfnamefont {S.}~\bibnamefont {Polla}}, \bibinfo {author} {\bibfnamefont
  {W.~J.}\ \bibnamefont {Huggins}}, \bibinfo {author} {\bibfnamefont
  {O.}~\bibnamefont {Oumarou}}, \bibinfo {author} {\bibfnamefont
  {K.}~\bibnamefont {Kechedzhi}}, \bibinfo {author} {\bibfnamefont
  {D.}~\bibnamefont {Abanin}}, \bibinfo {author} {\bibfnamefont
  {R.}~\bibnamefont {Acharya}}, \bibinfo {author} {\bibfnamefont
  {I.}~\bibnamefont {Aleiner}}, \bibinfo {author} {\bibfnamefont
  {R.}~\bibnamefont {Allen}}, \bibinfo {author} {\bibfnamefont {T.~I.}\
  \bibnamefont {Andersen}}, \bibinfo {author} {\bibfnamefont {K.}~\bibnamefont
  {Anderson}}, \bibinfo {author} {\bibfnamefont {M.}~\bibnamefont {Ansmann}},
  \bibinfo {author} {\bibfnamefont {F.}~\bibnamefont {Arute}}, \bibinfo
  {author} {\bibfnamefont {K.}~\bibnamefont {Arya}}, \bibinfo {author}
  {\bibfnamefont {A.}~\bibnamefont {Asfaw}}, \bibinfo {author} {\bibfnamefont
  {J.}~\bibnamefont {Atalaya}}, \bibinfo {author} {\bibfnamefont
  {D.}~\bibnamefont {Bacon}}, \bibinfo {author} {\bibfnamefont {J.~C.}\
  \bibnamefont {Bardin}}, \bibinfo {author} {\bibfnamefont {A.}~\bibnamefont
  {Bengtsson}}, \bibinfo {author} {\bibfnamefont {S.}~\bibnamefont {Boixo}},
  \bibinfo {author} {\bibfnamefont {G.}~\bibnamefont {Bortoli}}, \bibinfo
  {author} {\bibfnamefont {A.}~\bibnamefont {Bourassa}}, \bibinfo {author}
  {\bibfnamefont {J.}~\bibnamefont {Bovaird}}, \bibinfo {author} {\bibfnamefont
  {L.}~\bibnamefont {Brill}}, \bibinfo {author} {\bibfnamefont
  {M.}~\bibnamefont {Broughton}}, \bibinfo {author} {\bibfnamefont
  {B.}~\bibnamefont {Buckley}}, \bibinfo {author} {\bibfnamefont {D.~A.}\
  \bibnamefont {Buell}}, \bibinfo {author} {\bibfnamefont {T.}~\bibnamefont
  {Burger}}, \bibinfo {author} {\bibfnamefont {B.}~\bibnamefont {Burkett}},
  \bibinfo {author} {\bibfnamefont {N.}~\bibnamefont {Bushnell}}, \bibinfo
  {author} {\bibfnamefont {J.}~\bibnamefont {Campero}}, \bibinfo {author}
  {\bibfnamefont {Y.}~\bibnamefont {Chen}}, \bibinfo {author} {\bibfnamefont
  {Z.}~\bibnamefont {Chen}}, \bibinfo {author} {\bibfnamefont {B.}~\bibnamefont
  {Chiaro}}, \bibinfo {author} {\bibfnamefont {D.}~\bibnamefont {Chik}},
  \bibinfo {author} {\bibfnamefont {J.}~\bibnamefont {Cogan}}, \bibinfo
  {author} {\bibfnamefont {R.}~\bibnamefont {Collins}}, \bibinfo {author}
  {\bibfnamefont {P.}~\bibnamefont {Conner}}, \bibinfo {author} {\bibfnamefont
  {W.}~\bibnamefont {Courtney}}, \bibinfo {author} {\bibfnamefont {A.~L.}\
  \bibnamefont {Crook}}, \bibinfo {author} {\bibfnamefont {B.}~\bibnamefont
  {Curtin}}, \bibinfo {author} {\bibfnamefont {D.~M.}\ \bibnamefont {Debroy}},
  \bibinfo {author} {\bibfnamefont {S.}~\bibnamefont {Demura}}, \bibinfo
  {author} {\bibfnamefont {I.}~\bibnamefont {Drozdov}}, \bibinfo {author}
  {\bibfnamefont {A.}~\bibnamefont {Dunsworth}}, \bibinfo {author}
  {\bibfnamefont {C.}~\bibnamefont {Erickson}}, \bibinfo {author}
  {\bibfnamefont {L.}~\bibnamefont {Faoro}}, \bibinfo {author} {\bibfnamefont
  {E.}~\bibnamefont {Farhi}}, \bibinfo {author} {\bibfnamefont
  {R.}~\bibnamefont {Fatemi}}, \bibinfo {author} {\bibfnamefont {V.~S.}\
  \bibnamefont {Ferreira}}, \bibinfo {author} {\bibfnamefont {L.~F.}\
  \bibnamefont {Burgos}}, \bibinfo {author} {\bibfnamefont {E.}~\bibnamefont
  {Forati}}, \bibinfo {author} {\bibfnamefont {A.~G.}\ \bibnamefont {Fowler}},
  \bibinfo {author} {\bibfnamefont {B.}~\bibnamefont {Foxen}}, \bibinfo
  {author} {\bibfnamefont {W.}~\bibnamefont {Giang}}, \bibinfo {author}
  {\bibfnamefont {C.}~\bibnamefont {Gidney}}, \bibinfo {author} {\bibfnamefont
  {D.}~\bibnamefont {Gilboa}}, \bibinfo {author} {\bibfnamefont
  {M.}~\bibnamefont {Giustina}}, \bibinfo {author} {\bibfnamefont
  {R.}~\bibnamefont {Gosula}}, \bibinfo {author} {\bibfnamefont {A.~G.}\
  \bibnamefont {Dau}}, \bibinfo {author} {\bibfnamefont {J.~A.}\ \bibnamefont
  {Gross}}, \bibinfo {author} {\bibfnamefont {S.}~\bibnamefont {Habegger}},
  \bibinfo {author} {\bibfnamefont {M.~C.}\ \bibnamefont {Hamilton}}, \bibinfo
  {author} {\bibfnamefont {M.}~\bibnamefont {Hansen}}, \bibinfo {author}
  {\bibfnamefont {M.~P.}\ \bibnamefont {Harrigan}}, \bibinfo {author}
  {\bibfnamefont {S.~D.}\ \bibnamefont {Harrington}}, \bibinfo {author}
  {\bibfnamefont {P.}~\bibnamefont {Heu}}, \bibinfo {author} {\bibfnamefont
  {J.}~\bibnamefont {Hilton}}, \bibinfo {author} {\bibfnamefont {M.~R.}\
  \bibnamefont {Hoffmann}}, \bibinfo {author} {\bibfnamefont {S.}~\bibnamefont
  {Hong}}, \bibinfo {author} {\bibfnamefont {T.}~\bibnamefont {Huang}},
  \bibinfo {author} {\bibfnamefont {A.}~\bibnamefont {Huff}}, \bibinfo {author}
  {\bibfnamefont {L.~B.}\ \bibnamefont {Ioffe}}, \bibinfo {author}
  {\bibfnamefont {S.~V.}\ \bibnamefont {Isakov}}, \bibinfo {author}
  {\bibfnamefont {J.}~\bibnamefont {Iveland}}, \bibinfo {author} {\bibfnamefont
  {E.}~\bibnamefont {Jeffrey}}, \bibinfo {author} {\bibfnamefont
  {Z.}~\bibnamefont {Jiang}}, \bibinfo {author} {\bibfnamefont
  {C.}~\bibnamefont {Jones}}, \bibinfo {author} {\bibfnamefont
  {P.}~\bibnamefont {Juhas}}, \bibinfo {author} {\bibfnamefont
  {D.}~\bibnamefont {Kafri}}, \bibinfo {author} {\bibfnamefont
  {J.}~\bibnamefont {Kelly}}, \bibinfo {author} {\bibfnamefont
  {T.}~\bibnamefont {Khattar}}, \bibinfo {author} {\bibfnamefont
  {M.}~\bibnamefont {Khezri}}, \bibinfo {author} {\bibfnamefont
  {M.}~\bibnamefont {Kieferov{\'a}}}, \bibinfo {author} {\bibfnamefont
  {S.}~\bibnamefont {Kim}}, \bibinfo {author} {\bibfnamefont {P.~V.}\
  \bibnamefont {Klimov}}, \bibinfo {author} {\bibfnamefont {A.~R.}\
  \bibnamefont {Klots}}, \bibinfo {author} {\bibfnamefont {R.}~\bibnamefont
  {Kothari}}, \bibinfo {author} {\bibfnamefont {A.~N.}\ \bibnamefont
  {Korotkov}}, \bibinfo {author} {\bibfnamefont {F.}~\bibnamefont {Kostritsa}},
  \bibinfo {author} {\bibfnamefont {J.~M.}\ \bibnamefont {Kreikebaum}},
  \bibinfo {author} {\bibfnamefont {D.}~\bibnamefont {Landhuis}}, \bibinfo
  {author} {\bibfnamefont {P.}~\bibnamefont {Laptev}}, \bibinfo {author}
  {\bibfnamefont {K.}~\bibnamefont {Lau}}, \bibinfo {author} {\bibfnamefont
  {L.}~\bibnamefont {Laws}}, \bibinfo {author} {\bibfnamefont {J.}~\bibnamefont
  {Lee}}, \bibinfo {author} {\bibfnamefont {K.}~\bibnamefont {Lee}}, \bibinfo
  {author} {\bibfnamefont {B.~J.}\ \bibnamefont {Lester}}, \bibinfo {author}
  {\bibfnamefont {A.~T.}\ \bibnamefont {Lill}}, \bibinfo {author}
  {\bibfnamefont {W.}~\bibnamefont {Liu}}, \bibinfo {author} {\bibfnamefont
  {W.~P.}\ \bibnamefont {Livingston}}, \bibinfo {author} {\bibfnamefont
  {A.}~\bibnamefont {Locharla}}, \bibinfo {author} {\bibfnamefont
  {E.}~\bibnamefont {Lucero}}, \bibinfo {author} {\bibfnamefont {F.~D.}\
  \bibnamefont {Malone}}, \bibinfo {author} {\bibfnamefont {S.}~\bibnamefont
  {Mandra}}, \bibinfo {author} {\bibfnamefont {O.}~\bibnamefont {Martin}},
  \bibinfo {author} {\bibfnamefont {S.}~\bibnamefont {Martin}}, \bibinfo
  {author} {\bibfnamefont {J.~R.}\ \bibnamefont {McClean}}, \bibinfo {author}
  {\bibfnamefont {T.}~\bibnamefont {McCourt}}, \bibinfo {author} {\bibfnamefont
  {M.}~\bibnamefont {McEwen}}, \bibinfo {author} {\bibfnamefont
  {A.}~\bibnamefont {Megrant}}, \bibinfo {author} {\bibfnamefont
  {X.}~\bibnamefont {Mi}}, \bibinfo {author} {\bibfnamefont {A.}~\bibnamefont
  {Mieszala}}, \bibinfo {author} {\bibfnamefont {K.~C.}\ \bibnamefont {Miao}},
  \bibinfo {author} {\bibfnamefont {M.}~\bibnamefont {Mohseni}}, \bibinfo
  {author} {\bibfnamefont {S.}~\bibnamefont {Montazeri}}, \bibinfo {author}
  {\bibfnamefont {A.}~\bibnamefont {Morvan}}, \bibinfo {author} {\bibfnamefont
  {R.}~\bibnamefont {Movassagh}}, \bibinfo {author} {\bibfnamefont
  {W.}~\bibnamefont {Mruczkiewicz}}, \bibinfo {author} {\bibfnamefont
  {O.}~\bibnamefont {Naaman}}, \bibinfo {author} {\bibfnamefont
  {M.}~\bibnamefont {Neeley}}, \bibinfo {author} {\bibfnamefont
  {C.}~\bibnamefont {Neill}}, \bibinfo {author} {\bibfnamefont
  {A.}~\bibnamefont {Nersisyan}}, \bibinfo {author} {\bibfnamefont
  {H.}~\bibnamefont {Neven}}, \bibinfo {author} {\bibfnamefont
  {M.}~\bibnamefont {Newman}}, \bibinfo {author} {\bibfnamefont {J.~H.}\
  \bibnamefont {Ng}}, \bibinfo {author} {\bibfnamefont {A.}~\bibnamefont
  {Nguyen}}, \bibinfo {author} {\bibfnamefont {M.}~\bibnamefont {Nguyen}},
  \bibinfo {author} {\bibfnamefont {M.~Y.}\ \bibnamefont {Niu}}, \bibinfo
  {author} {\bibfnamefont {S.}~\bibnamefont {Omonije}}, \bibinfo {author}
  {\bibfnamefont {A.}~\bibnamefont {Opremcak}}, \bibinfo {author}
  {\bibfnamefont {A.}~\bibnamefont {Petukhov}}, \bibinfo {author}
  {\bibfnamefont {R.}~\bibnamefont {Potter}}, \bibinfo {author} {\bibfnamefont
  {L.~P.}\ \bibnamefont {Pryadko}}, \bibinfo {author} {\bibfnamefont
  {C.}~\bibnamefont {Quintana}}, \bibinfo {author} {\bibfnamefont
  {C.}~\bibnamefont {Rocque}}, \bibinfo {author} {\bibfnamefont
  {P.}~\bibnamefont {Roushan}}, \bibinfo {author} {\bibfnamefont
  {N.}~\bibnamefont {Saei}}, \bibinfo {author} {\bibfnamefont {D.}~\bibnamefont
  {Sank}}, \bibinfo {author} {\bibfnamefont {K.}~\bibnamefont
  {Sankaragomathi}}, \bibinfo {author} {\bibfnamefont {K.~J.}\ \bibnamefont
  {Satzinger}}, \bibinfo {author} {\bibfnamefont {H.~F.}\ \bibnamefont
  {Schurkus}}, \bibinfo {author} {\bibfnamefont {C.}~\bibnamefont {Schuster}},
  \bibinfo {author} {\bibfnamefont {M.~J.}\ \bibnamefont {Shearn}}, \bibinfo
  {author} {\bibfnamefont {A.}~\bibnamefont {Shorter}}, \bibinfo {author}
  {\bibfnamefont {N.}~\bibnamefont {Shutty}}, \bibinfo {author} {\bibfnamefont
  {V.}~\bibnamefont {Shvarts}}, \bibinfo {author} {\bibfnamefont
  {J.}~\bibnamefont {Skruzny}}, \bibinfo {author} {\bibfnamefont
  {V.}~\bibnamefont {Smelyanskiy}}, \bibinfo {author} {\bibfnamefont {W.~C.}\
  \bibnamefont {Smith}}, \bibinfo {author} {\bibfnamefont {R.}~\bibnamefont
  {Somma}}, \bibinfo {author} {\bibfnamefont {G.}~\bibnamefont {Sterling}},
  \bibinfo {author} {\bibfnamefont {D.}~\bibnamefont {Strain}}, \bibinfo
  {author} {\bibfnamefont {M.}~\bibnamefont {Szalay}}, \bibinfo {author}
  {\bibfnamefont {D.}~\bibnamefont {Thor}}, \bibinfo {author} {\bibfnamefont
  {A.}~\bibnamefont {Torres}}, \bibinfo {author} {\bibfnamefont
  {G.}~\bibnamefont {Vidal}}, \bibinfo {author} {\bibfnamefont
  {B.}~\bibnamefont {Villalonga}}, \bibinfo {author} {\bibfnamefont {C.~V.}\
  \bibnamefont {Heidweiller}}, \bibinfo {author} {\bibfnamefont
  {T.}~\bibnamefont {White}}, \bibinfo {author} {\bibfnamefont {B.~W.~K.}\
  \bibnamefont {Woo}}, \bibinfo {author} {\bibfnamefont {C.}~\bibnamefont
  {Xing}}, \bibinfo {author} {\bibfnamefont {Z.~J.}\ \bibnamefont {Yao}},
  \bibinfo {author} {\bibfnamefont {P.}~\bibnamefont {Yeh}}, \bibinfo {author}
  {\bibfnamefont {J.}~\bibnamefont {Yoo}}, \bibinfo {author} {\bibfnamefont
  {G.}~\bibnamefont {Young}}, \bibinfo {author} {\bibfnamefont
  {A.}~\bibnamefont {Zalcman}}, \bibinfo {author} {\bibfnamefont
  {Y.}~\bibnamefont {Zhang}}, \bibinfo {author} {\bibfnamefont
  {N.}~\bibnamefont {Zhu}}, \bibinfo {author} {\bibfnamefont {N.}~\bibnamefont
  {Zobrist}}, \bibinfo {author} {\bibfnamefont {C.}~\bibnamefont {Gogolin}},
  \bibinfo {author} {\bibfnamefont {R.}~\bibnamefont {Babbush}},\ and\ \bibinfo
  {author} {\bibfnamefont {N.~C.}\ \bibnamefont {Rubin}},\ }\href
  {https://doi.org/10.48550/arXiv.2210.10799} {\bibinfo {title}
  {Purification-based quantum error mitigation of pair-correlated electron
  simulations}} (\bibinfo {year} {2022}),\ \Eprint
  {https://arxiv.org/abs/2210.10799} {arxiv:2210.10799} \BibitemShut {NoStop}%
\bibitem [{\citenamefont {Tazhigulov}\ \emph {et~al.}(2022)\citenamefont
  {Tazhigulov}, \citenamefont {Sun}, \citenamefont {Haghshenas}, \citenamefont
  {Zhai}, \citenamefont {Tan}, \citenamefont {Rubin}, \citenamefont {Babbush},
  \citenamefont {Minnich},\ and\ \citenamefont
  {Chan}}]{tazhigulovSimulatingModelsChallenging2022}%
  \BibitemOpen
  \bibfield  {author} {\bibinfo {author} {\bibfnamefont {R.~N.}\ \bibnamefont
  {Tazhigulov}}, \bibinfo {author} {\bibfnamefont {S.-N.}\ \bibnamefont {Sun}},
  \bibinfo {author} {\bibfnamefont {R.}~\bibnamefont {Haghshenas}}, \bibinfo
  {author} {\bibfnamefont {H.}~\bibnamefont {Zhai}}, \bibinfo {author}
  {\bibfnamefont {A.~T.}\ \bibnamefont {Tan}}, \bibinfo {author} {\bibfnamefont
  {N.~C.}\ \bibnamefont {Rubin}}, \bibinfo {author} {\bibfnamefont
  {R.}~\bibnamefont {Babbush}}, \bibinfo {author} {\bibfnamefont {A.~J.}\
  \bibnamefont {Minnich}},\ and\ \bibinfo {author} {\bibfnamefont {G.~K.-L.}\
  \bibnamefont {Chan}},\ }\bibfield  {title} {\bibinfo {title} {Simulating
  {{Models}} of {{Challenging Correlated Molecules}} and {{Materials}} on the
  {{Sycamore Quantum Processor}}},\ }\href
  {https://doi.org/10.1103/PRXQuantum.3.040318} {\bibfield  {journal} {\bibinfo
   {journal} {PRX Quantum}\ }\textbf {\bibinfo {volume} {3}},\ \bibinfo {pages}
  {040318} (\bibinfo {year} {2022})}\BibitemShut {NoStop}%
\bibitem [{\citenamefont {Elfving}\ \emph {et~al.}(2020)\citenamefont
  {Elfving}, \citenamefont {Broer}, \citenamefont {Webber}, \citenamefont
  {Gavartin}, \citenamefont {Halls}, \citenamefont {Lorton},\ and\
  \citenamefont {Bochevarov}}]{elfvingHowWillQuantum2020}%
  \BibitemOpen
  \bibfield  {author} {\bibinfo {author} {\bibfnamefont {V.~E.}\ \bibnamefont
  {Elfving}}, \bibinfo {author} {\bibfnamefont {B.~W.}\ \bibnamefont {Broer}},
  \bibinfo {author} {\bibfnamefont {M.}~\bibnamefont {Webber}}, \bibinfo
  {author} {\bibfnamefont {J.}~\bibnamefont {Gavartin}}, \bibinfo {author}
  {\bibfnamefont {M.~D.}\ \bibnamefont {Halls}}, \bibinfo {author}
  {\bibfnamefont {K.~P.}\ \bibnamefont {Lorton}},\ and\ \bibinfo {author}
  {\bibfnamefont {A.}~\bibnamefont {Bochevarov}},\ }\href
  {https://doi.org/10.48550/arXiv.2009.12472} {\bibinfo {title} {How will
  quantum computers provide an industrially relevant computational advantage in
  quantum chemistry?}} (\bibinfo {year} {2020}),\ \Eprint
  {https://arxiv.org/abs/2009.12472} {arxiv:2009.12472} \BibitemShut {NoStop}%
\bibitem [{\citenamefont {Nagy}\ and\ \citenamefont
  {Jensen}(2017)}]{nagyBasisSetsQuantum2017}%
  \BibitemOpen
  \bibfield  {author} {\bibinfo {author} {\bibfnamefont {B.}~\bibnamefont
  {Nagy}}\ and\ \bibinfo {author} {\bibfnamefont {F.}~\bibnamefont {Jensen}},\
  }\bibfield  {title} {\bibinfo {title} {Basis {{Sets}} in {{Quantum
  Chemistry}}},\ }in\ \href {https://doi.org/10.1002/9781119356059.ch3} {\emph
  {\bibinfo {booktitle} {Reviews in {{Computational Chemistry}}}}}\ (\bibinfo
  {publisher} {{John Wiley \& Sons, Ltd}},\ \bibinfo {year} {2017})\
  Chap.~\bibinfo {chapter} {3}, pp.\ \bibinfo {pages} {93--149}\BibitemShut
  {NoStop}%
\bibitem [{\citenamefont {Romero}\ \emph {et~al.}(2018)\citenamefont {Romero},
  \citenamefont {Babbush}, \citenamefont {McClean}, \citenamefont {Hempel},
  \citenamefont {Love},\ and\ \citenamefont
  {{Aspuru-Guzik}}}]{romeroStrategiesQuantumComputing2018}%
  \BibitemOpen
  \bibfield  {author} {\bibinfo {author} {\bibfnamefont {J.}~\bibnamefont
  {Romero}}, \bibinfo {author} {\bibfnamefont {R.}~\bibnamefont {Babbush}},
  \bibinfo {author} {\bibfnamefont {J.~R.}\ \bibnamefont {McClean}}, \bibinfo
  {author} {\bibfnamefont {C.}~\bibnamefont {Hempel}}, \bibinfo {author}
  {\bibfnamefont {P.~J.}\ \bibnamefont {Love}},\ and\ \bibinfo {author}
  {\bibfnamefont {A.}~\bibnamefont {{Aspuru-Guzik}}},\ }\bibfield  {title}
  {\bibinfo {title} {Strategies for quantum computing molecular energies using
  the unitary coupled cluster ansatz},\ }\href
  {https://doi.org/10.1088/2058-9565/aad3e4} {\bibfield  {journal} {\bibinfo
  {journal} {Quantum Sci. Technol.}\ }\textbf {\bibinfo {volume} {4}},\
  \bibinfo {pages} {014008} (\bibinfo {year} {2018})}\BibitemShut {NoStop}%
\bibitem [{\citenamefont {Arrazola}\ \emph {et~al.}(2022)\citenamefont
  {Arrazola}, \citenamefont {Di~Matteo}, \citenamefont {Quesada}, \citenamefont
  {Jahangiri}, \citenamefont {Delgado},\ and\ \citenamefont
  {Killoran}}]{arrazolaUniversalQuantumCircuits2022}%
  \BibitemOpen
  \bibfield  {author} {\bibinfo {author} {\bibfnamefont {J.~M.}\ \bibnamefont
  {Arrazola}}, \bibinfo {author} {\bibfnamefont {O.}~\bibnamefont {Di~Matteo}},
  \bibinfo {author} {\bibfnamefont {N.}~\bibnamefont {Quesada}}, \bibinfo
  {author} {\bibfnamefont {S.}~\bibnamefont {Jahangiri}}, \bibinfo {author}
  {\bibfnamefont {A.}~\bibnamefont {Delgado}},\ and\ \bibinfo {author}
  {\bibfnamefont {N.}~\bibnamefont {Killoran}},\ }\bibfield  {title} {\bibinfo
  {title} {Universal quantum circuits for quantum chemistry},\ }\href
  {https://doi.org/10.22331/q-2022-06-20-742} {\bibfield  {journal} {\bibinfo
  {journal} {Quantum}\ }\textbf {\bibinfo {volume} {6}},\ \bibinfo {pages}
  {742} (\bibinfo {year} {2022})},\ \Eprint {https://arxiv.org/abs/2106.13839}
  {arxiv:2106.13839} \BibitemShut {NoStop}%
\bibitem [{\citenamefont {Anselmetti}\ \emph {et~al.}(2021)\citenamefont
  {Anselmetti}, \citenamefont {Wierichs}, \citenamefont {Gogolin},\ and\
  \citenamefont
  {Parrish}}]{anselmettiLocalExpressiveQuantumnumberpreserving2021}%
  \BibitemOpen
  \bibfield  {author} {\bibinfo {author} {\bibfnamefont {G.-L.~R.}\
  \bibnamefont {Anselmetti}}, \bibinfo {author} {\bibfnamefont
  {D.}~\bibnamefont {Wierichs}}, \bibinfo {author} {\bibfnamefont
  {C.}~\bibnamefont {Gogolin}},\ and\ \bibinfo {author} {\bibfnamefont {R.~M.}\
  \bibnamefont {Parrish}},\ }\bibfield  {title} {\bibinfo {title} {Local,
  expressive, quantum-number-preserving {{VQE}} ans\"atze for fermionic
  systems},\ }\href {https://doi.org/10.1088/1367-2630/ac2cb3} {\bibfield
  {journal} {\bibinfo  {journal} {New J. Phys.}\ }\textbf {\bibinfo {volume}
  {23}},\ \bibinfo {pages} {113010} (\bibinfo {year} {2021})}\BibitemShut
  {NoStop}%
\bibitem [{\citenamefont {Yordanov}\ \emph {et~al.}(2020)\citenamefont
  {Yordanov}, \citenamefont {{Arvidsson-Shukur}},\ and\ \citenamefont
  {Barnes}}]{yordanovEfficientQuantumCircuits2020}%
  \BibitemOpen
  \bibfield  {author} {\bibinfo {author} {\bibfnamefont {Y.~S.}\ \bibnamefont
  {Yordanov}}, \bibinfo {author} {\bibfnamefont {D.~R.~M.}\ \bibnamefont
  {{Arvidsson-Shukur}}},\ and\ \bibinfo {author} {\bibfnamefont {C.~H.~W.}\
  \bibnamefont {Barnes}},\ }\bibfield  {title} {\bibinfo {title} {Efficient
  quantum circuits for quantum computational chemistry},\ }\href
  {https://doi.org/10.1103/PhysRevA.102.062612} {\bibfield  {journal} {\bibinfo
   {journal} {Phys. Rev. A}\ }\textbf {\bibinfo {volume} {102}},\ \bibinfo
  {pages} {062612} (\bibinfo {year} {2020})}\BibitemShut {NoStop}%
\bibitem [{\citenamefont {Magoulas}\ and\ \citenamefont
  {Evangelista}(2023)}]{magoulasCNOTEfficientCircuitsArbitrary2023}%
  \BibitemOpen
  \bibfield  {author} {\bibinfo {author} {\bibfnamefont {I.}~\bibnamefont
  {Magoulas}}\ and\ \bibinfo {author} {\bibfnamefont {F.~A.}\ \bibnamefont
  {Evangelista}},\ }\bibfield  {title} {\bibinfo {title} {{{CNOT-Efficient
  Circuits}} for {{Arbitrary Rank Many-Body Fermionic}} and {{Qubit
  Excitations}}},\ }\href {https://doi.org/10.1021/acs.jctc.2c01016} {\bibfield
   {journal} {\bibinfo  {journal} {J. Chem. Theory Comput.}\ }\textbf {\bibinfo
  {volume} {19}},\ \bibinfo {pages} {822} (\bibinfo {year} {2023})}\BibitemShut
  {NoStop}%
\bibitem [{\citenamefont {Chee}\ \emph {et~al.}(2022)\citenamefont {Chee},
  \citenamefont {Mak}, \citenamefont {Leykam}, \citenamefont {Barkoutsos},\
  and\ \citenamefont {Angelakis}}]{cheeComputingElectronicCorrelation2022}%
  \BibitemOpen
  \bibfield  {author} {\bibinfo {author} {\bibfnamefont {C.~H.}\ \bibnamefont
  {Chee}}, \bibinfo {author} {\bibfnamefont {A.~M.}\ \bibnamefont {Mak}},
  \bibinfo {author} {\bibfnamefont {D.}~\bibnamefont {Leykam}}, \bibinfo
  {author} {\bibfnamefont {P.~K.}\ \bibnamefont {Barkoutsos}},\ and\ \bibinfo
  {author} {\bibfnamefont {D.~G.}\ \bibnamefont {Angelakis}},\ }\href
  {https://doi.org/10.48550/arXiv.2207.03949} {\bibinfo {title} {Computing
  {{Electronic Correlation Energies}} using {{Linear Depth Quantum Circuits}}}}
  (\bibinfo {year} {2022}),\ \Eprint {https://arxiv.org/abs/2207.03949}
  {arxiv:2207.03949} \BibitemShut {NoStop}%
\bibitem [{\citenamefont {Kerenidis}\ and\ \citenamefont
  {Prakash}(2022)}]{kerenidisQuantumMachineLearning2022}%
  \BibitemOpen
  \bibfield  {author} {\bibinfo {author} {\bibfnamefont {I.}~\bibnamefont
  {Kerenidis}}\ and\ \bibinfo {author} {\bibfnamefont {A.}~\bibnamefont
  {Prakash}},\ }\href {https://doi.org/10.48550/arXiv.2202.00054} {\bibinfo
  {title} {Quantum machine learning with subspace states}} (\bibinfo {year}
  {2022}),\ \Eprint {https://arxiv.org/abs/2202.00054} {arxiv:2202.00054}
  \BibitemShut {NoStop}%
\bibitem [{\citenamefont
  {Stewart}(1982)}]{stewartComputingTheCSDecomposition1982}%
  \BibitemOpen
  \bibfield  {author} {\bibinfo {author} {\bibfnamefont {G.~W.}\ \bibnamefont
  {Stewart}},\ }\bibfield  {title} {\bibinfo {title} {Computing {{theCS}}
  decomposition of a partitioned orthonormal matrix},\ }\href
  {https://doi.org/10.1007/BF01396447} {\bibfield  {journal} {\bibinfo
  {journal} {Numer. Math.}\ }\textbf {\bibinfo {volume} {40}},\ \bibinfo
  {pages} {297} (\bibinfo {year} {1982})}\BibitemShut {NoStop}%
\bibitem [{\citenamefont {Gawlik}\ \emph {et~al.}(2018)\citenamefont {Gawlik},
  \citenamefont {Nakatsukasa},\ and\ \citenamefont
  {Sutton}}]{gawlikBackwardStableAlgorithm2018}%
  \BibitemOpen
  \bibfield  {author} {\bibinfo {author} {\bibfnamefont {E.~S.}\ \bibnamefont
  {Gawlik}}, \bibinfo {author} {\bibfnamefont {Y.}~\bibnamefont
  {Nakatsukasa}},\ and\ \bibinfo {author} {\bibfnamefont {B.~D.}\ \bibnamefont
  {Sutton}},\ }\bibfield  {title} {\bibinfo {title} {A {{Backward Stable
  Algorithm}} for {{Computing}} the {{CS Decomposition}} via the {{Polar
  Decomposition}}},\ }\href {https://doi.org/10.1137/18M1182747} {\bibfield
  {journal} {\bibinfo  {journal} {SIAM J. Matrix Anal. Appl.}\ }\textbf
  {\bibinfo {volume} {39}},\ \bibinfo {pages} {1448} (\bibinfo {year}
  {2018})}\BibitemShut {NoStop}%
\bibitem [{\citenamefont {Johri}\ \emph {et~al.}(2021)\citenamefont {Johri},
  \citenamefont {Debnath}, \citenamefont {Mocherla}, \citenamefont {Singk},
  \citenamefont {Prakash}, \citenamefont {Kim},\ and\ \citenamefont
  {Kerenidis}}]{johriNearestCentroidClassification2021}%
  \BibitemOpen
  \bibfield  {author} {\bibinfo {author} {\bibfnamefont {S.}~\bibnamefont
  {Johri}}, \bibinfo {author} {\bibfnamefont {S.}~\bibnamefont {Debnath}},
  \bibinfo {author} {\bibfnamefont {A.}~\bibnamefont {Mocherla}}, \bibinfo
  {author} {\bibfnamefont {A.}~\bibnamefont {Singk}}, \bibinfo {author}
  {\bibfnamefont {A.}~\bibnamefont {Prakash}}, \bibinfo {author} {\bibfnamefont
  {J.}~\bibnamefont {Kim}},\ and\ \bibinfo {author} {\bibfnamefont
  {I.}~\bibnamefont {Kerenidis}},\ }\bibfield  {title} {\bibinfo {title}
  {Nearest centroid classification on a trapped ion quantum computer},\ }\href
  {https://doi.org/10.1038/s41534-021-00456-5} {\bibfield  {journal} {\bibinfo
  {journal} {npj Quantum Inf}\ }\textbf {\bibinfo {volume} {7}},\ \bibinfo
  {pages} {1} (\bibinfo {year} {2021})}\BibitemShut {NoStop}%
\bibitem [{\citenamefont {Wan}\ \emph {et~al.}(2022)\citenamefont {Wan},
  \citenamefont {Huggins}, \citenamefont {Lee},\ and\ \citenamefont
  {Babbush}}]{wanMatchgateShadowsFermionic2022}%
  \BibitemOpen
  \bibfield  {author} {\bibinfo {author} {\bibfnamefont {K.}~\bibnamefont
  {Wan}}, \bibinfo {author} {\bibfnamefont {W.~J.}\ \bibnamefont {Huggins}},
  \bibinfo {author} {\bibfnamefont {J.}~\bibnamefont {Lee}},\ and\ \bibinfo
  {author} {\bibfnamefont {R.}~\bibnamefont {Babbush}},\ }\href
  {https://doi.org/10.48550/arXiv.2207.13723} {\bibinfo {title} {Matchgate
  {{Shadows}} for {{Fermionic Quantum Simulation}}}} (\bibinfo {year} {2022}),\
  \Eprint {https://arxiv.org/abs/2207.13723} {arxiv:2207.13723} \BibitemShut
  {NoStop}%
\bibitem [{\citenamefont {Ortiz}\ \emph {et~al.}(2001)\citenamefont {Ortiz},
  \citenamefont {Gubernatis}, \citenamefont {Knill},\ and\ \citenamefont
  {Laflamme}}]{ortizQuantumAlgorithmsFermionic2001}%
  \BibitemOpen
  \bibfield  {author} {\bibinfo {author} {\bibfnamefont {G.}~\bibnamefont
  {Ortiz}}, \bibinfo {author} {\bibfnamefont {J.~E.}\ \bibnamefont
  {Gubernatis}}, \bibinfo {author} {\bibfnamefont {E.}~\bibnamefont {Knill}},\
  and\ \bibinfo {author} {\bibfnamefont {R.}~\bibnamefont {Laflamme}},\
  }\bibfield  {title} {\bibinfo {title} {Quantum algorithms for fermionic
  simulations},\ }\href {https://doi.org/10.1103/PhysRevA.64.022319} {\bibfield
   {journal} {\bibinfo  {journal} {Phys. Rev. A}\ }\textbf {\bibinfo {volume}
  {64}},\ \bibinfo {pages} {022319} (\bibinfo {year} {2001})}\BibitemShut
  {NoStop}%
\bibitem [{\citenamefont {Jiang}\ \emph {et~al.}(2018)\citenamefont {Jiang},
  \citenamefont {Sung}, \citenamefont {Kechedzhi}, \citenamefont
  {Smelyanskiy},\ and\ \citenamefont
  {Boixo}}]{jiangQuantumAlgorithmsSimulate2018}%
  \BibitemOpen
  \bibfield  {author} {\bibinfo {author} {\bibfnamefont {Z.}~\bibnamefont
  {Jiang}}, \bibinfo {author} {\bibfnamefont {K.~J.}\ \bibnamefont {Sung}},
  \bibinfo {author} {\bibfnamefont {K.}~\bibnamefont {Kechedzhi}}, \bibinfo
  {author} {\bibfnamefont {V.~N.}\ \bibnamefont {Smelyanskiy}},\ and\ \bibinfo
  {author} {\bibfnamefont {S.}~\bibnamefont {Boixo}},\ }\bibfield  {title}
  {\bibinfo {title} {Quantum {{Algorithms}} to {{Simulate Many-Body Physics}}
  of {{Correlated Fermions}}},\ }\href
  {https://doi.org/10.1103/PhysRevApplied.9.044036} {\bibfield  {journal}
  {\bibinfo  {journal} {Phys. Rev. Appl.}\ }\textbf {\bibinfo {volume} {9}},\
  \bibinfo {pages} {044036} (\bibinfo {year} {2018})}\BibitemShut {NoStop}%
\bibitem [{\citenamefont
  {Vourdas}(2018)}]{vourdasExteriorCalculusFermionic2018}%
  \BibitemOpen
  \bibfield  {author} {\bibinfo {author} {\bibfnamefont {A.}~\bibnamefont
  {Vourdas}},\ }\bibfield  {title} {\bibinfo {title} {Exterior calculus and
  fermionic quantum computation},\ }\href
  {https://doi.org/10.1088/1751-8121/aae2c4} {\bibfield  {journal} {\bibinfo
  {journal} {J. Phys. A: Math. Theor.}\ }\textbf {\bibinfo {volume} {51}},\
  \bibinfo {pages} {445301} (\bibinfo {year} {2018})}\BibitemShut {NoStop}%
\bibitem [{\citenamefont {Kerenidis}\ \emph {et~al.}(2021)\citenamefont
  {Kerenidis}, \citenamefont {Landman},\ and\ \citenamefont
  {Mathur}}]{kerenidisClassicalQuantumAlgorithms2021}%
  \BibitemOpen
  \bibfield  {author} {\bibinfo {author} {\bibfnamefont {I.}~\bibnamefont
  {Kerenidis}}, \bibinfo {author} {\bibfnamefont {J.}~\bibnamefont {Landman}},\
  and\ \bibinfo {author} {\bibfnamefont {N.}~\bibnamefont {Mathur}},\ }\href
  {https://doi.org/10.48550/arXiv.2106.07198} {\bibinfo {title} {Classical and
  {{Quantum Algorithms}} for {{Orthogonal Neural Networks}}}} (\bibinfo {year}
  {2021}),\ \Eprint {https://arxiv.org/abs/2106.07198} {arxiv:2106.07198}
  \BibitemShut {NoStop}%
\bibitem [{\citenamefont {Whitfield}\ \emph {et~al.}(2011)\citenamefont
  {Whitfield}, \citenamefont {Biamonte},\ and\ \citenamefont
  {{Aspuru-Guzik}}}]{whitfieldSimulationElectronicStructure2011}%
  \BibitemOpen
  \bibfield  {author} {\bibinfo {author} {\bibfnamefont {J.~D.}\ \bibnamefont
  {Whitfield}}, \bibinfo {author} {\bibfnamefont {J.}~\bibnamefont
  {Biamonte}},\ and\ \bibinfo {author} {\bibfnamefont {A.}~\bibnamefont
  {{Aspuru-Guzik}}},\ }\bibfield  {title} {\bibinfo {title} {Simulation of
  electronic structure {{Hamiltonians}} using quantum computers},\ }\href
  {https://doi.org/10.1080/00268976.2011.552441} {\bibfield  {journal}
  {\bibinfo  {journal} {Molecular Physics}\ }\textbf {\bibinfo {volume}
  {109}},\ \bibinfo {pages} {735} (\bibinfo {year} {2011})}\BibitemShut
  {NoStop}%
\bibitem [{\citenamefont {Kjaergaard}\ \emph {et~al.}(2020)\citenamefont
  {Kjaergaard}, \citenamefont {Schwartz}, \citenamefont {Braum{\"u}ller},
  \citenamefont {Krantz}, \citenamefont {Wang}, \citenamefont {Gustavsson},\
  and\ \citenamefont {Oliver}}]{kjaergaardSuperconductingQubitsCurrent2020}%
  \BibitemOpen
  \bibfield  {author} {\bibinfo {author} {\bibfnamefont {M.}~\bibnamefont
  {Kjaergaard}}, \bibinfo {author} {\bibfnamefont {M.~E.}\ \bibnamefont
  {Schwartz}}, \bibinfo {author} {\bibfnamefont {J.}~\bibnamefont
  {Braum{\"u}ller}}, \bibinfo {author} {\bibfnamefont {P.}~\bibnamefont
  {Krantz}}, \bibinfo {author} {\bibfnamefont {J.~I.-J.}\ \bibnamefont {Wang}},
  \bibinfo {author} {\bibfnamefont {S.}~\bibnamefont {Gustavsson}},\ and\
  \bibinfo {author} {\bibfnamefont {W.~D.}\ \bibnamefont {Oliver}},\ }\bibfield
   {title} {\bibinfo {title} {Superconducting {{Qubits}}: {{Current State}} of
  {{Play}}},\ }\href {https://doi.org/10.1146/annurev-conmatphys-031119-050605}
  {\bibfield  {journal} {\bibinfo  {journal} {Annu. Rev. Condens. Matter
  Phys.}\ }\textbf {\bibinfo {volume} {11}},\ \bibinfo {pages} {369} (\bibinfo
  {year} {2020})}\BibitemShut {NoStop}%
\bibitem [{\citenamefont {Bruzewicz}\ \emph {et~al.}(2019)\citenamefont
  {Bruzewicz}, \citenamefont {Chiaverini}, \citenamefont {McConnell},\ and\
  \citenamefont {Sage}}]{bruzewiczTrappedionQuantumComputing2019}%
  \BibitemOpen
  \bibfield  {author} {\bibinfo {author} {\bibfnamefont {C.~D.}\ \bibnamefont
  {Bruzewicz}}, \bibinfo {author} {\bibfnamefont {J.}~\bibnamefont
  {Chiaverini}}, \bibinfo {author} {\bibfnamefont {R.}~\bibnamefont
  {McConnell}},\ and\ \bibinfo {author} {\bibfnamefont {J.~M.}\ \bibnamefont
  {Sage}},\ }\bibfield  {title} {\bibinfo {title} {Trapped-ion quantum
  computing: {{Progress}} and challenges},\ }\href
  {https://doi.org/10.1063/1.5088164} {\bibfield  {journal} {\bibinfo
  {journal} {Applied Physics Reviews}\ }\textbf {\bibinfo {volume} {6}},\
  \bibinfo {pages} {021314} (\bibinfo {year} {2019})}\BibitemShut {NoStop}%
\bibitem [{\citenamefont {Martin}(1996)}]{martinInitioTotalAtomization1996}%
  \BibitemOpen
  \bibfield  {author} {\bibinfo {author} {\bibfnamefont {J.~M.~L.}\
  \bibnamefont {Martin}},\ }\bibfield  {title} {\bibinfo {title} {Ab initio
  total atomization energies of small molecules \textemdash{} towards the basis
  set limit},\ }\href {https://doi.org/10.1016/0009-2614(96)00898-6} {\bibfield
   {journal} {\bibinfo  {journal} {Chemical Physics Letters}\ }\textbf
  {\bibinfo {volume} {259}},\ \bibinfo {pages} {669} (\bibinfo {year}
  {1996})}\BibitemShut {NoStop}%
\bibitem [{\citenamefont {Halkier}\ \emph {et~al.}(1999)\citenamefont
  {Halkier}, \citenamefont {Helgaker}, \citenamefont {J{\o}rgensen},
  \citenamefont {Klopper},\ and\ \citenamefont
  {Olsen}}]{halkierBasissetConvergenceEnergy1999}%
  \BibitemOpen
  \bibfield  {author} {\bibinfo {author} {\bibfnamefont {A.}~\bibnamefont
  {Halkier}}, \bibinfo {author} {\bibfnamefont {T.}~\bibnamefont {Helgaker}},
  \bibinfo {author} {\bibfnamefont {P.}~\bibnamefont {J{\o}rgensen}}, \bibinfo
  {author} {\bibfnamefont {W.}~\bibnamefont {Klopper}},\ and\ \bibinfo {author}
  {\bibfnamefont {J.}~\bibnamefont {Olsen}},\ }\bibfield  {title} {\bibinfo
  {title} {Basis-set convergence of the energy in molecular
  {{Hartree}}\textendash{{Fock}} calculations},\ }\href
  {https://doi.org/10.1016/S0009-2614(99)00179-7} {\bibfield  {journal}
  {\bibinfo  {journal} {Chemical Physics Letters}\ }\textbf {\bibinfo {volume}
  {302}},\ \bibinfo {pages} {437} (\bibinfo {year} {1999})}\BibitemShut
  {NoStop}%
\bibitem [{\citenamefont {Spackman}\ and\ \citenamefont
  {Karton}(2015)}]{spackmanEstimatingCCSDBasisset2015}%
  \BibitemOpen
  \bibfield  {author} {\bibinfo {author} {\bibfnamefont {P.~R.}\ \bibnamefont
  {Spackman}}\ and\ \bibinfo {author} {\bibfnamefont {A.}~\bibnamefont
  {Karton}},\ }\bibfield  {title} {\bibinfo {title} {Estimating the {{CCSD}}
  basis-set limit energy from small basis sets: Basis-set extrapolations vs
  additivity schemes},\ }\href {https://doi.org/10.1063/1.4921697} {\bibfield
  {journal} {\bibinfo  {journal} {AIP Advances}\ }\textbf {\bibinfo {volume}
  {5}},\ \bibinfo {pages} {057148} (\bibinfo {year} {2015})}\BibitemShut
  {NoStop}%
\bibitem [{\citenamefont {Plascencia}\ \emph {et~al.}(2017)\citenamefont
  {Plascencia}, \citenamefont {Wang},\ and\ \citenamefont
  {Wilson}}]{plascenciaImportanceLigandBasis2017}%
  \BibitemOpen
  \bibfield  {author} {\bibinfo {author} {\bibfnamefont {C.}~\bibnamefont
  {Plascencia}}, \bibinfo {author} {\bibfnamefont {J.}~\bibnamefont {Wang}},\
  and\ \bibinfo {author} {\bibfnamefont {A.~K.}\ \bibnamefont {Wilson}},\
  }\bibfield  {title} {\bibinfo {title} {Importance of the ligand basis set in
  ab initio thermochemical calculations of transition metal species},\ }\href
  {https://doi.org/10.1016/j.cplett.2017.08.003} {\bibfield  {journal}
  {\bibinfo  {journal} {Chemical Physics Letters}\ }\textbf {\bibinfo {volume}
  {685}},\ \bibinfo {pages} {496} (\bibinfo {year} {2017})}\BibitemShut
  {NoStop}%
\bibitem [{\citenamefont {Virtanen}\ \emph {et~al.}(2020)\citenamefont
  {Virtanen}, \citenamefont {Gommers}, \citenamefont {Oliphant}, \citenamefont
  {Haberland}, \citenamefont {Reddy}, \citenamefont {Cournapeau}, \citenamefont
  {Burovski}, \citenamefont {Peterson}, \citenamefont {Weckesser},
  \citenamefont {Bright}, \citenamefont {{van der Walt}}, \citenamefont
  {Brett}, \citenamefont {Wilson}, \citenamefont {Millman}, \citenamefont
  {Mayorov}, \citenamefont {Nelson}, \citenamefont {Jones}, \citenamefont
  {Kern}, \citenamefont {Larson}, \citenamefont {Carey}, \citenamefont {Polat},
  \citenamefont {Feng}, \citenamefont {Moore}, \citenamefont {VanderPlas},
  \citenamefont {Laxalde}, \citenamefont {Perktold}, \citenamefont {Cimrman},
  \citenamefont {Henriksen}, \citenamefont {Quintero}, \citenamefont {Harris},
  \citenamefont {Archibald}, \citenamefont {Ribeiro}, \citenamefont
  {Pedregosa},\ and\ \citenamefont {{van
  Mulbregt}}}]{virtanenSciPyFundamentalAlgorithms2020}%
  \BibitemOpen
  \bibfield  {author} {\bibinfo {author} {\bibfnamefont {P.}~\bibnamefont
  {Virtanen}}, \bibinfo {author} {\bibfnamefont {R.}~\bibnamefont {Gommers}},
  \bibinfo {author} {\bibfnamefont {T.~E.}\ \bibnamefont {Oliphant}}, \bibinfo
  {author} {\bibfnamefont {M.}~\bibnamefont {Haberland}}, \bibinfo {author}
  {\bibfnamefont {T.}~\bibnamefont {Reddy}}, \bibinfo {author} {\bibfnamefont
  {D.}~\bibnamefont {Cournapeau}}, \bibinfo {author} {\bibfnamefont
  {E.}~\bibnamefont {Burovski}}, \bibinfo {author} {\bibfnamefont
  {P.}~\bibnamefont {Peterson}}, \bibinfo {author} {\bibfnamefont
  {W.}~\bibnamefont {Weckesser}}, \bibinfo {author} {\bibfnamefont
  {J.}~\bibnamefont {Bright}}, \bibinfo {author} {\bibfnamefont {S.~J.}\
  \bibnamefont {{van der Walt}}}, \bibinfo {author} {\bibfnamefont
  {M.}~\bibnamefont {Brett}}, \bibinfo {author} {\bibfnamefont
  {J.}~\bibnamefont {Wilson}}, \bibinfo {author} {\bibfnamefont {K.~J.}\
  \bibnamefont {Millman}}, \bibinfo {author} {\bibfnamefont {N.}~\bibnamefont
  {Mayorov}}, \bibinfo {author} {\bibfnamefont {A.~R.~J.}\ \bibnamefont
  {Nelson}}, \bibinfo {author} {\bibfnamefont {E.}~\bibnamefont {Jones}},
  \bibinfo {author} {\bibfnamefont {R.}~\bibnamefont {Kern}}, \bibinfo {author}
  {\bibfnamefont {E.}~\bibnamefont {Larson}}, \bibinfo {author} {\bibfnamefont
  {C.~J.}\ \bibnamefont {Carey}}, \bibinfo {author} {\bibfnamefont
  {{\.I}.}~\bibnamefont {Polat}}, \bibinfo {author} {\bibfnamefont
  {Y.}~\bibnamefont {Feng}}, \bibinfo {author} {\bibfnamefont {E.~W.}\
  \bibnamefont {Moore}}, \bibinfo {author} {\bibfnamefont {J.}~\bibnamefont
  {VanderPlas}}, \bibinfo {author} {\bibfnamefont {D.}~\bibnamefont {Laxalde}},
  \bibinfo {author} {\bibfnamefont {J.}~\bibnamefont {Perktold}}, \bibinfo
  {author} {\bibfnamefont {R.}~\bibnamefont {Cimrman}}, \bibinfo {author}
  {\bibfnamefont {I.}~\bibnamefont {Henriksen}}, \bibinfo {author}
  {\bibfnamefont {E.~A.}\ \bibnamefont {Quintero}}, \bibinfo {author}
  {\bibfnamefont {C.~R.}\ \bibnamefont {Harris}}, \bibinfo {author}
  {\bibfnamefont {A.~M.}\ \bibnamefont {Archibald}}, \bibinfo {author}
  {\bibfnamefont {A.~H.}\ \bibnamefont {Ribeiro}}, \bibinfo {author}
  {\bibfnamefont {F.}~\bibnamefont {Pedregosa}},\ and\ \bibinfo {author}
  {\bibfnamefont {P.}~\bibnamefont {{van Mulbregt}}},\ }\bibfield  {title}
  {\bibinfo {title} {{{SciPy}} 1.0: Fundamental algorithms for scientific
  computing in {{Python}}},\ }\href {https://doi.org/10.1038/s41592-019-0686-2}
  {\bibfield  {journal} {\bibinfo  {journal} {Nat. Methods}\ }\textbf {\bibinfo
  {volume} {17}},\ \bibinfo {pages} {261} (\bibinfo {year} {2020})}\BibitemShut
  {NoStop}%
\bibitem [{\citenamefont {Sun}\ \emph {et~al.}(2018)\citenamefont {Sun},
  \citenamefont {Berkelbach}, \citenamefont {Blunt}, \citenamefont {Booth},
  \citenamefont {Guo}, \citenamefont {Li}, \citenamefont {Liu}, \citenamefont
  {McClain}, \citenamefont {Sayfutyarova}, \citenamefont {Sharma},
  \citenamefont {Wouters},\ and\ \citenamefont
  {Chan}}]{sunPySCFPythonbasedSimulations2018}%
  \BibitemOpen
  \bibfield  {author} {\bibinfo {author} {\bibfnamefont {Q.}~\bibnamefont
  {Sun}}, \bibinfo {author} {\bibfnamefont {T.~C.}\ \bibnamefont {Berkelbach}},
  \bibinfo {author} {\bibfnamefont {N.~S.}\ \bibnamefont {Blunt}}, \bibinfo
  {author} {\bibfnamefont {G.~H.}\ \bibnamefont {Booth}}, \bibinfo {author}
  {\bibfnamefont {S.}~\bibnamefont {Guo}}, \bibinfo {author} {\bibfnamefont
  {Z.}~\bibnamefont {Li}}, \bibinfo {author} {\bibfnamefont {J.}~\bibnamefont
  {Liu}}, \bibinfo {author} {\bibfnamefont {J.~D.}\ \bibnamefont {McClain}},
  \bibinfo {author} {\bibfnamefont {E.~R.}\ \bibnamefont {Sayfutyarova}},
  \bibinfo {author} {\bibfnamefont {S.}~\bibnamefont {Sharma}}, \bibinfo
  {author} {\bibfnamefont {S.}~\bibnamefont {Wouters}},\ and\ \bibinfo {author}
  {\bibfnamefont {G.~K.-L.}\ \bibnamefont {Chan}},\ }\bibfield  {title}
  {\bibinfo {title} {{{PySCF}}: The {{Python-based}} simulations of chemistry
  framework},\ }\href {https://doi.org/10.1002/wcms.1340} {\bibfield  {journal}
  {\bibinfo  {journal} {WIREs Computational Molecular Science}\ }\textbf
  {\bibinfo {volume} {8}},\ \bibinfo {pages} {e1340} (\bibinfo {year}
  {2018})}\BibitemShut {NoStop}%
\bibitem [{\citenamefont {Bergholm}\ \emph {et~al.}(2022)\citenamefont
  {Bergholm}, \citenamefont {Izaac}, \citenamefont {Schuld}, \citenamefont
  {Gogolin}, \citenamefont {Ahmed}, \citenamefont {Ajith}, \citenamefont
  {Alam}, \citenamefont {{Alonso-Linaje}}, \citenamefont {AkashNarayanan},
  \citenamefont {Asadi}, \citenamefont {Arrazola}, \citenamefont {Azad},
  \citenamefont {Banning}, \citenamefont {Blank}, \citenamefont {Bromley},
  \citenamefont {Cordier}, \citenamefont {Ceroni}, \citenamefont {Delgado},
  \citenamefont {Di~Matteo}, \citenamefont {Dusko}, \citenamefont {Garg},
  \citenamefont {Guala}, \citenamefont {Hayes}, \citenamefont {Hill},
  \citenamefont {Ijaz}, \citenamefont {Isacsson}, \citenamefont {Ittah},
  \citenamefont {Jahangiri}, \citenamefont {Jain}, \citenamefont {Jiang},
  \citenamefont {Khandelwal}, \citenamefont {Kottmann}, \citenamefont {Lang},
  \citenamefont {Lee}, \citenamefont {Loke}, \citenamefont {Lowe},
  \citenamefont {McKiernan}, \citenamefont {Meyer}, \citenamefont
  {{Monta{\~n}ez-Barrera}}, \citenamefont {Moyard}, \citenamefont {Niu},
  \citenamefont {O'Riordan}, \citenamefont {Oud}, \citenamefont {Panigrahi},
  \citenamefont {Park}, \citenamefont {Polatajko}, \citenamefont {Quesada},
  \citenamefont {Roberts}, \citenamefont {S{\'a}}, \citenamefont {Schoch},
  \citenamefont {Shi}, \citenamefont {Shu}, \citenamefont {Sim}, \citenamefont
  {Singh}, \citenamefont {Strandberg}, \citenamefont {Soni}, \citenamefont
  {Sz{\'a}va}, \citenamefont {Thabet}, \citenamefont {{Vargas-Hern{\'a}ndez}},
  \citenamefont {Vincent}, \citenamefont {Vitucci}, \citenamefont {Weber},
  \citenamefont {Wierichs}, \citenamefont {Wiersema}, \citenamefont {Willmann},
  \citenamefont {Wong}, \citenamefont {Zhang},\ and\ \citenamefont
  {Killoran}}]{bergholmPennyLaneAutomaticDifferentiation2022}%
  \BibitemOpen
  \bibfield  {author} {\bibinfo {author} {\bibfnamefont {V.}~\bibnamefont
  {Bergholm}}, \bibinfo {author} {\bibfnamefont {J.}~\bibnamefont {Izaac}},
  \bibinfo {author} {\bibfnamefont {M.}~\bibnamefont {Schuld}}, \bibinfo
  {author} {\bibfnamefont {C.}~\bibnamefont {Gogolin}}, \bibinfo {author}
  {\bibfnamefont {S.}~\bibnamefont {Ahmed}}, \bibinfo {author} {\bibfnamefont
  {V.}~\bibnamefont {Ajith}}, \bibinfo {author} {\bibfnamefont {M.~S.}\
  \bibnamefont {Alam}}, \bibinfo {author} {\bibfnamefont {G.}~\bibnamefont
  {{Alonso-Linaje}}}, \bibinfo {author} {\bibfnamefont {B.}~\bibnamefont
  {AkashNarayanan}}, \bibinfo {author} {\bibfnamefont {A.}~\bibnamefont
  {Asadi}}, \bibinfo {author} {\bibfnamefont {J.~M.}\ \bibnamefont {Arrazola}},
  \bibinfo {author} {\bibfnamefont {U.}~\bibnamefont {Azad}}, \bibinfo {author}
  {\bibfnamefont {S.}~\bibnamefont {Banning}}, \bibinfo {author} {\bibfnamefont
  {C.}~\bibnamefont {Blank}}, \bibinfo {author} {\bibfnamefont {T.~R.}\
  \bibnamefont {Bromley}}, \bibinfo {author} {\bibfnamefont {B.~A.}\
  \bibnamefont {Cordier}}, \bibinfo {author} {\bibfnamefont {J.}~\bibnamefont
  {Ceroni}}, \bibinfo {author} {\bibfnamefont {A.}~\bibnamefont {Delgado}},
  \bibinfo {author} {\bibfnamefont {O.}~\bibnamefont {Di~Matteo}}, \bibinfo
  {author} {\bibfnamefont {A.}~\bibnamefont {Dusko}}, \bibinfo {author}
  {\bibfnamefont {T.}~\bibnamefont {Garg}}, \bibinfo {author} {\bibfnamefont
  {D.}~\bibnamefont {Guala}}, \bibinfo {author} {\bibfnamefont
  {A.}~\bibnamefont {Hayes}}, \bibinfo {author} {\bibfnamefont
  {R.}~\bibnamefont {Hill}}, \bibinfo {author} {\bibfnamefont {A.}~\bibnamefont
  {Ijaz}}, \bibinfo {author} {\bibfnamefont {T.}~\bibnamefont {Isacsson}},
  \bibinfo {author} {\bibfnamefont {D.}~\bibnamefont {Ittah}}, \bibinfo
  {author} {\bibfnamefont {S.}~\bibnamefont {Jahangiri}}, \bibinfo {author}
  {\bibfnamefont {P.}~\bibnamefont {Jain}}, \bibinfo {author} {\bibfnamefont
  {E.}~\bibnamefont {Jiang}}, \bibinfo {author} {\bibfnamefont
  {A.}~\bibnamefont {Khandelwal}}, \bibinfo {author} {\bibfnamefont
  {K.}~\bibnamefont {Kottmann}}, \bibinfo {author} {\bibfnamefont {R.~A.}\
  \bibnamefont {Lang}}, \bibinfo {author} {\bibfnamefont {C.}~\bibnamefont
  {Lee}}, \bibinfo {author} {\bibfnamefont {T.}~\bibnamefont {Loke}}, \bibinfo
  {author} {\bibfnamefont {A.}~\bibnamefont {Lowe}}, \bibinfo {author}
  {\bibfnamefont {K.}~\bibnamefont {McKiernan}}, \bibinfo {author}
  {\bibfnamefont {J.~J.}\ \bibnamefont {Meyer}}, \bibinfo {author}
  {\bibfnamefont {J.~A.}\ \bibnamefont {{Monta{\~n}ez-Barrera}}}, \bibinfo
  {author} {\bibfnamefont {R.}~\bibnamefont {Moyard}}, \bibinfo {author}
  {\bibfnamefont {Z.}~\bibnamefont {Niu}}, \bibinfo {author} {\bibfnamefont
  {L.~J.}\ \bibnamefont {O'Riordan}}, \bibinfo {author} {\bibfnamefont
  {S.}~\bibnamefont {Oud}}, \bibinfo {author} {\bibfnamefont {A.}~\bibnamefont
  {Panigrahi}}, \bibinfo {author} {\bibfnamefont {C.-Y.}\ \bibnamefont {Park}},
  \bibinfo {author} {\bibfnamefont {D.}~\bibnamefont {Polatajko}}, \bibinfo
  {author} {\bibfnamefont {N.}~\bibnamefont {Quesada}}, \bibinfo {author}
  {\bibfnamefont {C.}~\bibnamefont {Roberts}}, \bibinfo {author} {\bibfnamefont
  {N.}~\bibnamefont {S{\'a}}}, \bibinfo {author} {\bibfnamefont
  {I.}~\bibnamefont {Schoch}}, \bibinfo {author} {\bibfnamefont
  {B.}~\bibnamefont {Shi}}, \bibinfo {author} {\bibfnamefont {S.}~\bibnamefont
  {Shu}}, \bibinfo {author} {\bibfnamefont {S.}~\bibnamefont {Sim}}, \bibinfo
  {author} {\bibfnamefont {A.}~\bibnamefont {Singh}}, \bibinfo {author}
  {\bibfnamefont {I.}~\bibnamefont {Strandberg}}, \bibinfo {author}
  {\bibfnamefont {J.}~\bibnamefont {Soni}}, \bibinfo {author} {\bibfnamefont
  {A.}~\bibnamefont {Sz{\'a}va}}, \bibinfo {author} {\bibfnamefont
  {S.}~\bibnamefont {Thabet}}, \bibinfo {author} {\bibfnamefont {R.~A.}\
  \bibnamefont {{Vargas-Hern{\'a}ndez}}}, \bibinfo {author} {\bibfnamefont
  {T.}~\bibnamefont {Vincent}}, \bibinfo {author} {\bibfnamefont
  {N.}~\bibnamefont {Vitucci}}, \bibinfo {author} {\bibfnamefont
  {M.}~\bibnamefont {Weber}}, \bibinfo {author} {\bibfnamefont
  {D.}~\bibnamefont {Wierichs}}, \bibinfo {author} {\bibfnamefont
  {R.}~\bibnamefont {Wiersema}}, \bibinfo {author} {\bibfnamefont
  {M.}~\bibnamefont {Willmann}}, \bibinfo {author} {\bibfnamefont
  {V.}~\bibnamefont {Wong}}, \bibinfo {author} {\bibfnamefont {S.}~\bibnamefont
  {Zhang}},\ and\ \bibinfo {author} {\bibfnamefont {N.}~\bibnamefont
  {Killoran}},\ }\href {https://doi.org/10.48550/arXiv.1811.04968} {\bibinfo
  {title} {{{PennyLane}}: {{Automatic}} differentiation of hybrid
  quantum-classical computations}} (\bibinfo {year} {2022}),\ \Eprint
  {https://arxiv.org/abs/1811.04968} {arxiv:1811.04968} \BibitemShut {NoStop}%
\end{thebibliography}%
\end{document}